\documentclass[sigconf, nonacm]{acmart}

\usepackage[ruled,vlined,linesnumbered]{algorithm2e} 
\usepackage{multirow,graphicx,float,subfig,caption,fancyhdr,xcolor,enumitem,colortbl}
\usepackage{amsfonts}	
\usepackage{epstopdf}
\usepackage{bbding}
\usepackage{array}
\usepackage{hyperref}
\usepackage{pifont}
\usepackage[most]{tcolorbox}
\usepackage{tikz}
\usepackage[utf8]{inputenc}
\usepackage{lipsum} % for dummy text
\setcopyright{none} 
\SetKwProg{Fn}{Function}{:}{}
\SetKwProg{Proc}{Procedure}{:}{}
\SetKwFunction{upd}{Update}
\SetKwFunction{ub}{UpperBound}
\SetKwFunction{setub}{SetUB}
\SetKwFunction{heu}{Heuristic}
\SetKwFunction{br}{Branch}
\SetKwFunction{bdd}{BranchBDD}
\SetKwFunction{bkpo}{BranchImp}
\SetKwFunction{greedy}{GreedySearch}
\SetKwFunction{dmbp}{DMBP}
\SetKwFunction{fastbb}{FastBB}
\SetKwFunction{cpc}{CMBS-CSS}
\SetKwFunction{bppivot}{BPPivot}
\SetKw{Break}{break}
\SetKw{Continue}{continue}
\SetKw{And}{and}
\SetKw{Or}{or}
\SetKw{In}{in}
\SetKw{To}{to}

\DeclareSubrefFormat{parens}{#1(#2)}
\newtheoremstyle{proofs}{3pt}{3pt}{}{}{\scshape}{.}{ }{}

\theoremstyle{proofs}

\newtheorem{theorem}{\hspace{1em}Theorem}%[theorem]
\newtheorem{lemma}{\hspace{1em}Lemma}%[lemma]

\theoremstyle{definition}
\newtheorem*{problem}{Problem statement}
\newtheorem*{detail}{Implementation details}

\theoremstyle{remark}

\let\savedbaselinestretch\baselinestretch

\setlist[itemize]{leftmargin=2em, topsep=0pt, partopsep=0pt, parsep=0pt, itemsep=0pt}
\setlist[enumerate]{leftmargin=2em, topsep=0pt, partopsep=0pt, parsep=0pt, itemsep=0pt}

\makeatletter
\g@addto@macro\normalsize{%
	\setlength\abovedisplayskip{2pt plus 1pt minus 1pt} % 从4pt降到2pt
	\setlength\belowdisplayskip{2pt plus 1pt minus 1pt}
	\setlength\abovedisplayshortskip{0pt}
	\setlength\belowdisplayshortskip{1pt}
}
\makeatother

\begin{document}
	
\author{
	{Donghang Cui\texorpdfstring{$^\dagger$}\,,
		Rong-Hua Li\texorpdfstring{$^\dagger$}\,,
		Qiangqiang Dai\texorpdfstring{$^\dagger$}\,, 
		%Hongchao Qin\texorpdfstring{$^\dagger$}\,,
		Guoren Wang\texorpdfstring{$^\dagger$},}
}
\affiliation{
	{\texorpdfstring{$^\dagger$} BBeijing Institute of Technology}
	\country{China}
}
\email{
	{cuidonghang@bit.edu.cn, 
		lironghuabit@126.com, 
		qiangd66@gmail.com, 
		%qhc.neu@gmail.com, 
		wanggrbit@126.com}
}
	
%\title{Efficient Maximum Biplex Search: A Deletion-based Approach}
\title{Revisiting Maximum $k$-Biplex Search Through $k$-Bounded-Degree Deletion}

\begin{abstract}
	Biplex, as a relaxation of the biclique model, has emerged as an important cohesive subgraph model for bipartite graph analysis. The maximum $k$-biplex search problem aims to identify the $k$-biplex with maximum number of edges and has been widely applied in various real-world applications, including community detection, online recommendation, and fraud detection. However, the problem is NP-hard, and existing exact algorithms remain inefficient on large-scale bipartite graphs with large values of $k$ (e.g., $k\geq 3$).
	In this paper, we revisit the maximum $k$-biplex search problem from a complementary perspective. We reveal a novel structural duality: finding a maximum $k$-biplex in a bipartite graph is equivalent to finding a minimal $k$-bounded-degree deletion in its complement graph. Based on this observation, we propose a novel deletion-based algorithm for the maximum $k$-biplex search problem. We theoretically prove that the proposed algorithm achieves a worst-case time complexity of $O^*(\gamma_k^n)$, where $\gamma_k<2$. Specifically, $\gamma_1=1.725$, $\gamma_2=1.856$, and $\gamma_3=1.928$.
	To further enhance practical efficiency, we develop several effective upper-bounding techniques and a heuristic strategy for obtaining high-quality initial solutions, which substantially reduce the search space. Extensive experiments on eight real-world bipartite graphs demonstrate the efficiency of our approach, which achieves up to four orders of magnitude speedups over state-of-the-art algorithms.

\end{abstract}
	
\maketitle

\section{Introduction} 

Bipartite graphs are fundamental data structures for modeling interactions between two distinct groups of entities, such as user-item interactions in e-commerce networks \cite{RN50,RN71}, author-publication relationships in academic networks \cite{RN34}, and gene-protein associations in computational biology networks \cite{RN60,RN38,RN40}. Identifying cohesive subgraphs is a core task in bipartite graph mining, since such subgraphs often reveal meaningful groups with strong cross-side associations. For example, in computational biology networks, cohesive subgraphs can capture groups of genes and proteins with strong expression associations, thereby helping identify functional modules or disease-related biological pathways \cite{RN40}.

As a classical cohesive subgraph model in bipartite graphs, the biclique has been widely studied and applied in many scenarios \cite{RN68,RN57,RN69}. A biclique is a complete bipartite subgraph, where every vertex on one side is connected to all vertices on the other side. The maximum biclique search problem aims to find a biclique containing the maximum number of edges, which has attracted considerable attention in recent years \cite{RN47,RN58,RN66,RN14,RN22,RN13}. However, real-world network data are inevitably affected by observation noise, measurement errors, and missing data \cite{RN70,RN130}. Due to the strict complete-connectivity requirement, the biclique is unable to represent cohesive communities that contain a small number of missing edges, which limits its practical application.

To address this limitation, the $k$-biplex model was introduced as a relaxation of the biclique and has attracted increasing attention in recent years \cite{RN72,RN75,RN66,RN74,RN73,RN71,RN152}. Formally, in a $k$-biplex of a bipartite graph, each vertex has at most $k$ non-neighbors. This relaxation allows limited missing edges while preserving its cohesion, and has been applied to many real-world tasks such as online recommendation \cite{RN41,RN42}, community detection \cite{RN66,RN161}, and fraud detection \cite{RN66,RN71}. In this paper, we study the problem of finding a $k$-biplex with the maximum number of edges, which is referred to as the maximum $k$-biplex search problem. This problem has been proven to be NP-hard for any positive integer $k$ \cite{RN152}, implying that no polynomial-time algorithm exists unless $\mathrm{P}=\mathrm{NP}$.

To solve the maximum $k$-biplex search problem, several exact algorithms have been developed in recent years \cite{RN73,RN71,RN152}. These algorithms mainly follow the branch-and-bound framework by deciding which vertices should be included in the target $k$-biplex, and they reduce the number of branches with carefully-designed branching and pruning strategies. However, these algorithms are mainly effective for small $k$ values, and their performance substantially degrades when $k$ becomes large, as shown in Table~\ref{tab:overall-runtime}. For example, on most real-world graphs, existing algorithms struggle to solve most test cases when $k\ge 3$. In particular, on the dataset \textit{Aol}, existing algorithms fail to handle any test cases even when $k\ge 2$.

Notably, several studies have proposed algorithms for maximal $k$-biplex enumeration \cite{RN74,RN66,RN72,RN75}. A natural way to apply these algorithms to the maximum $k$-biplex search problem is to enumerate all maximal $k$-biplexes and then select the one with the maximum number of edges. However, the number of maximal $k$-biplexes in a bipartite graph can be exponential with respect to the graph size. Directly applying these enumeration algorithms may generate a large number of maximal but non-maximum $k$-biplexes, thereby causing a large amount of unnecessary computation and making them inefficient in practice. Consequently, developing a theoretically and practically efficient algorithm for the maximum $k$-biplex search remains a challenging task.

\noindent\textbf{Contributions.} To address the aforementioned challenges, we propose a novel deletion-based algorithm for the maximum $k$-biplex search problem with a non-trivial worst-case time complexity. We also propose several optimization techniques to further improve the practical performance of the algorithm. The main contributions of this paper are as follows.

\noindent\underline{\emph{A novel deletion-based algorithm.}} We reveal the structural connection between maximum $k$-biplex search in the original bipartite graph and $k$-bounded-degree deletion in the complement graph. Based on this connection, we propose a deletion-based algorithm for maximum $k$-biplex search. Specifically, we first propose a novel deletion-based branching strategy, which decides which vertices to be deleted from the subgraph. We then develop several branch reduction techniques to further enhance the performance of the deletion-based branching strategy. Finally, we prove that our deletion-based algorithm achieves a non-trivial worst-case time complexity of $O^*(\gamma_k^n)$. When $k=1,2$ and $3$, we have $\gamma_1=1.725$, $\gamma_2=1.856$ and $\gamma_3=1.928$, respectively. To the best of our knowledge, our algorithm achieves the best known time complexity among all existing algorithms for maximum $k$-biplex search.

\noindent\underline{\emph{Efficient optimization techniques.}} To further improve the practical performance of the proposed deletion-based algorithm, we design several optimization techniques tailored to the deletion-based search framework. Specifically, we propose two efficient edge upper bounds to prune search branches that cannot generate a larger $k$-biplex. The first upper bound is derived from the newly added vertex and limits how many of its non-neighbors can still be kept; the second upper bound is derived from the total missing-edge allowance of the current partial solution on one side. We show that these upper bounds can be computed in linear time for each search instance. Finally, we develop a heuristic algorithm to quickly construct a large initial $k$-biplex in polynomial time, combining a greedy deletion stage with a localized expansion stage based on the distance property of large $k$-biplexes.

\noindent\underline{\emph{Comprehensive experiments.}} We conduct extensive experiments on 8 real-world bipartite graphs to evaluate the efficiency and scalability of our proposed algorithm. We compare our algorithm with two state-of-the-art maximum $k$-biplex search algorithms under various settings of $k$ and $\theta$. The experimental results demonstrate that our proposed algorithm substantially outperforms existing solutions, achieving up to four orders of magnitude speedups. For example, our algorithm solves all test cases on the graph \textit{Aol}, while the baseline algorithms fail to solve most test cases within 24 hours. More specifically, our algorithm finds a maximum $4$-biplex on \textit{Aol} with both side sizes larger than $17$ in only $6.14$ seconds, while the other two baseline algorithms cannot finish within 24 hours, demonstrating that our algorithm is up to four orders of magnitude faster than the state-of-the-art approaches.

\section{Preliminaries}

Let $G=(U,V,E)$ be a bipartite graph, where $U$ and $V$ are two disjoint sets of vertices, and $E\subseteq U\times V$ is the set of edges between vertices in $U$ and $V$. Let $n=\lvert U\rvert+\lvert V\rvert$ and $m=\lvert E\rvert$ be the number of vertices and edges in $G$, respectively. For every vertex $u\in U$ (resp. $v\in V$), let $N(u)=\{v\in V\ \vert\ (u,v)\in E\}$ (resp. $N(v)=\{u\in U\ \vert\ (u,v)\in E\}$) be the set of $u$'s (resp. $v$'s) neighbors in $G$, and let $\overline N(u)=V\setminus N(u)$ (resp. $\overline N(v)=U\setminus N(v)$) be the set of $u$'s (resp. $v$'s) non-neighbors in $G$. For every vertex $v\in U\cup V$, let $d(v)=\lvert N(v)\rvert$ and $\overline d(v)=\lvert \overline N(v)\rvert$ be the degree and non-degree of $v$ in $G$, respectively. Let $\delta=\max_{v\in U\cup V} d(v)$ and $\overline\delta=\max_{v\in U\cup V}\overline d(v)$ be the maximum degree and maximum non-degree of vertices in $G$, respectively.
Given a vertex subset $S=(U_S,V_S)$ of $G$, where $U_S\subseteq U$ and $V_S\subseteq V$, let $G[S]=(U_S,V_S,E_S)$ be the subgraph of $G$ induced by $S$, where $E_S=\{(u,v)\in E\ \vert\ u\in U_S\land v\in V_S\}$ is the set of edges between vertices in $U_S$ and $V_S$. For every vertex $v\in S$, let $N_S(v)=N(v)\cap S$ and $\overline N_S(v)=\overline N(v)\cap S$ be the set of $v$'s neighbors and non-neighbors in $G[S]$, respectively. Let $d_S(v)=\lvert N_S(v)\rvert$ and $\overline d_S(v)=\lvert \overline N_S(v)\rvert$ be the degree and non-degree of $v$ in $G[S]$, respectively. Let $\delta_S=\max_{v\in S} d_S(v)$ and $\overline\delta_S=\max_{v\in S}\overline d_S(v)$ be the maximum degree and maximum non-degree of vertices in $G[S]$, respectively.
Given a bipartite graph $G=(U,V,E)$, let $\overline G=(U,V,\overline E)$ be the complement graph of $G$, where $\overline E=U\times V\setminus E$ is the set of all missing edges in $G$. Given a vertex subset $S=(U_S,V_S)$, let $\overline G[S]=(U_S,V_S,\overline E_S)$ be the subgraph of $\overline G$ induced by $S$, i.e., the complement graph of $G[S]$, where $\overline E_S=U_S\times V_S\setminus E_S$ is the set of all missing edges in $G[S]$.
All the notations frequently used in this paper are presented in Table \ref{tab:not}. 

\begin{table}[!t]
	\caption{Notations frequently used in this paper}
	\label{tab:not}
	%\smaller[1]
	\resizebox{\columnwidth}{!}{\begin{tabular}{ll}
				\toprule
				Notations&Descriptions\\
				\midrule
				$G=(U,V,E)$&a bipartite graph\\
				$\overline G=(U,V,\overline E)$&the complement graph of $G$\\
				$S=(U_S,V_S)$&a vertex subset of $G$, with $U_S\subseteq U$ and $V_S\subseteq V$\\
				$G[S]=(U_S,V_S,E_S)$&a subgraph of $G$ induced by $S$\\
				$\overline E_S$&the set of non-edges in $G[S]$\\
				$N(v)$,\, $\overline N(v)$&the set of $v$'s neighbors and non-neighbors in $G$\\
				$d(v)$,\, $\overline d(v)$&the number of $v$'s neighbors and non-neighbors in $G$\\
				$N_S(v)$,\, $\overline N_S(v)$&the set of $v$'s neighbors and non-neighbors in $S$\\
				$d_S(v)$,\, $\overline d_S(v)$&the number of $v$'s neighbors and non-neighbors in $S$\\
				$\delta$,\, $\overline\delta$&the maximum degrees of vertices in $G$ and $\overline G$\\
				$\delta_S$,\, $\overline\delta_S$&the maximum degrees of vertices in $G[S]$ and $\overline G[S]$\\
				%$\theta$&the size threshold for MDB search problem\\
				%$S(v)$&the subset of $S$ with vertices located at $v$'s side (either $U(S)$ or $V_S$)\\ 
				\bottomrule
			\end{tabular}}
\end{table}

Formally, the $k$-biplex is defined as follows.

\begin{definition}[$k$-biplex]
	For a bipartite graph $G=(U,V,E)$, a $k$-biplex $G[S]$ is a subgraph of $G$ induced by $S=(U_S,V_S)$, where every vertex has at most $k$ non-neighbors, i.e., $\forall v\in S,\ \overline d_S(v)\le k$.
\end{definition}

A $k$-biplex of $G$ is called maximal if it cannot be included into another $k$-biplex of $G$, and is called maximum if it has the most edges among all maximal $k$-biplexes of $G$. For real-world bipartite graphs, we aim to find the maximum $k$-biplex to identify their most representative community. However, the maximum $k$-biplex of a bipartite graph may be disconnected, which is unsuitable for representing a meaningful community. To overcome this problem, we can constraint the number of vertices in each side of the target maximum $k$-biplex to ensure its connectivity, as presented in the following lemma \cite{RN71}.

\begin{lemma}
	A $k$-biplex $S=(U_S,V_S)$ is connected if $\lvert U_S\rvert\ge 2k+1$ and $\lvert V_S\rvert\ge 2k+1$.
\end{lemma}

Furthermore, a $k$-biplex with highly unbalanced number of vertices between the two sides (e.g. a 1-biplex where one side contains only 3 vertices while the other comprises thousands) has limited practical significance despite its connectivity. Therefore, we impose a threshold $\theta$ on the number of vertices for both sides. Formally, we define the problem investigated in this study as follows.

\begin{problem}[Maximum $k$-biplex search]
	Given a bipartite graph $G=(U,V,E)$ and two positive integers $\theta$ and $k$, where $\theta\ge 2k+1$, the maximum $k$-biplex search problem aims to find a $k$-biplex $S=(U_S,V_S,E_S)$ with the maximum number of edges in $G$ such that $\lvert U_S\rvert\ge \theta$ and $\lvert V_S\rvert\ge \theta$.
\end{problem}

\noindent\textbf{NP-hardness.} As established in \cite{RN71}, the maximum $k$-biplex search problem is NP-hard, which implies there is no polynomial algorithm for solving it unless $\mathrm{P}=\mathrm{NP}$. Consequently, it is essential to develop a maximum $k$-biplex search algorithm that is both theoretically and practically efficient.                 a

\subsection{Existing Solutions}
Several exact approaches have been proposed to solve the maximum $k$-biplex search problem \cite{RN71,RN152}. Below, we briefly review their main technical ideas and limitations.

Yu et al. \cite{RN71} studied the problem of finding $K$ maximal $k$-biplexes with the most edges, where $K=1$ corresponds to our problem. They proposed a branch-and-bound algorithm based on a symmetric Bron--Kerbosch branching strategy. Specifically, the algorithm chooses a pivot vertex $u$ with at least $k+1$ non-neighbors $v_1,v_2,\ldots,v_{k+1}$ in the current subgraph, and then generates subbranches in the following way: the first branch excludes $u$ from the subgraph, and the $i+1$-th branch excludes $v_i$ while adding $u,v_1,\ldots,v_{i-1}$ into the $k$-biplex. Since $u$ has at most $k$ non-neighbors in a $k$-biplex, at most $k+2$ subbranches will be generated. With this branching strategy, their algorithm achieves a worst-case time complexity $O^*(\alpha_k^n)$, where $\alpha_k<2$; when $k=1,2,$ and $3$, $\alpha_k=1.754,1.888,$ and $1.947$, respectively.
This branching strategy is effective when $k$ is small, but becomes less efficient when $k$ increases. This is because the number of subbranches generated by the symmetric-BK branching strategy depends heavily on $k$. When $k$ increases, the rapid increase in the number of branches results in an excessively large search space, thereby reducing search efficiency.

Pan et al. \cite{RN152} studied the maximum $k$-biplex search problem over large bipartite graphs and proposed a core-based graph reduction framework. Specifically, for every possible side-size pair $(x,y)$ of the target $k$-biplex, they compute the corresponding $(x,y)$-core, where vertices on the two sides have degree at least $y-k$ and $x-k$, respectively. The algorithm then examines every pair $(x,y)$ in decreasing order of the value $x\cdot y$. For the current pair $(x,y)$, it runs branch-and-bound search on the corresponding $(x,y)$-core to find a $k$-biplex $S=(U_S,V_S,E_S)$ with $|U_S|=x$ and $|V_S|=y$. Once such a solution is found, all remaining pairs $(x',y')$ with $x'\cdot y'\le |E_S|$ can be safely discarded, because no $k$-biplex with these side sizes can contain more than $x'\cdot y'$ edges. If no such $k$-biplex is found in the current core, the algorithm continues to the next pair.
However, the effectiveness of this core-based reduction decreases as $k$ becomes large. For a fixed side-size pair $(x,y)$, the degree requirements of the corresponding core are $y-k$ and $x-k$, respectively. Thus, a larger $k$ directly lowers these requirements and allows more vertices to remain in the reduced graph. As a result, the branch-and-bound search is still conducted on a large core. Moreover, since a larger $k$ allows more missing edges in the target $k$-biplex, the pruning rules become less restrictive, further increasing the search cost.

\section{A Novel Deletion-based Algorithm}
In this section, we propose a novel deletion-based algorithm for the maximum $k$-biplex search problem, which solves the problem from the complement graph: instead of deciding which vertices should be kept in the original graph, we decide which vertices should be deleted so that the remaining complement subgraph has bounded degree. We prove that our algorithm achieves a non-trivial worst-case time complexity of $O^*(\gamma_k^n)$, where $\gamma_k<2$. For example, when $k=1,2$ and $3$, we have $\gamma_1=1.725$, $\gamma_2=1.856$ and $\gamma_3=1.928$, respectively. To the best of our knowledge, our algorithm achieves the best known time complexity among all existing algorithms for maximum $k$-biplex search.
\subsection{Deletion-based Branching Framework} \label{sec:dbs}

We first introduce the concept of deletion in bipartite graphs. Given a bipartite graph $G=(U,V,E)$, the $k$-bounded-degree deletion of $G$ is defined as follows.

\begin{definition}[$k$-bounded-degree deletion]
	Given a bipartite graph $G=(U,V,E)$ and a positive integer $k$, a $k$-bounded-degree deletion of $G$ is a vertex subset $D\subseteq U\cup V$ such that after deleting all vertices in $D$ from $G$, the maximum degree of the remaining subgraph $G[(U\cup V)\setminus D]$ is at most $k$.
\end{definition}

For convenience of discussion, we refer to a $k$-bounded-degree deletion as a $k$-BDD in this paper. We say a $k$-BDD $D$ is minimal in $G$ if there is no other $k$-BDD $D'$ in $G$ satisfying that $D'\subset D$. 
We then discuss the relationship between $k$-biplex and $k$-BDD in bipartite graphs.
Given a $k$-biplex $S$ of $G$, let $D=(U\cup V)\setminus S$. Since every vertex in $S$ has at most $k$ non-neighbors in $G[S]$, the maximum degree of $\overline G[S]$ is at most $k$. Hence, $D$ is a $k$-BDD of $\overline G$. Conversely, if $D$ is a $k$-BDD of $\overline G$, then the remaining vertex set $(U\cup V)\setminus D$ induces a $k$-biplex in $G$.
However, this correspondence does not mean that the maximum $k$-biplex can be obtained by finding a $k$-BDD that leaves the fewest edges in the complement graph. For example, consider $k=1$ and $\theta=2$. Let $U=\{u_1,u_2,u_3\}$ and $V=\{v_1,v_2,v_3\}$, and let $G$ contain all edges except $(u_3,v_3)$. Then $G[U\cup V]$ is a $1$-biplex with eight edges, and thus it is the maximum $1$-biplex. In $\overline G$, this solution leaves one complement edge. In contrast, if we keep only $\{u_1,u_2\}\cup\{v_1,v_2\}$, the remaining complement graph has no edge, but the corresponding subgraph of $G$ has only four edges, and hence is not the maximum $k$-biplex of $G$. 
To overcome this limitation, we develop the new relationship between the maximum $k$-biplex of $G$ and the $k$-BDD of $\overline G$, which builds upon the following lemma.

\begin{lemma} \label{lem:dual}
	Given a bipartite graph $G=(U,V,E)$ and a maximum $k$-biplex $G[S]=(U_S,V_S,E_S)$ of $G$, $(U\cup V)\setminus S$ forms a minimal $k$-BDD of $\overline G$.
\end{lemma}

\begin{proof}
	We prove this lemma by contradiction. Let $D=(U\cup V)\setminus S$. If there exists a minimal $k$-BDD $D'$ in $\overline G$ satisfying $D'\subset D$, we have $S=(U\cup V)\setminus D\subset (U\cup V)\setminus D'$, which means $G[(U\cup V)\setminus D']$ forms a $k$-biplex that contains $G[S]$, contradicting the maximality. Therefore, $G[S]$ is not a maximum $k$-biplex of $G$, which contradicts the condition.
\end{proof}

Based on this lemma, we propose a new structural duality that a maximum $k$-biplex of $G$ always corresponds to a specific minimal $k$-BDD of $\overline G$, as shown in the following theorem.

\begin{theorem}[Structural Duality] \label{thm:dual}
	Let $G=(U,V,E)$ be a bipartite graph and $\overline G=(U,V,\overline E)$ be its complement. If a minimal $k$-BDD $D=(U_D,V_D)$ of $\overline G$ satisfies $\lvert U_D\rvert\le\lvert U\rvert-\theta$ and $\lvert V_D\rvert\le\lvert V\rvert-\theta$, and maximizes $\lvert U\setminus U_D\rvert\times\lvert V\setminus V_D\rvert-\lvert \overline E_{(U\cup V)\setminus D}\rvert$ among all minimal $k$-BDDs in $\overline G$ satisfying these two constraints, then $G[U\cup V\setminus D]$ is a maximum $k$-biplex of $G$.
\end{theorem}

\begin{proof}
	Let $S=(U\setminus U_D,V\setminus V_D)$. Since $D$ is a $k$-BDD of $\overline G$, the maximum degree of $\overline G[S]$ is at most $k$. Equivalently, every vertex in $G[S]$ has at most $k$ non-neighbors. Thus, $G[S]$ is a $k$-biplex. As $\lvert U_D\rvert\le\lvert U\rvert-\theta$ and $\lvert V_D\rvert\le\lvert V\rvert-\theta$, we have $\lvert U_S\rvert=\lvert U\setminus U_D\rvert\ge\theta$ and $\lvert V_S\rvert=\lvert V\setminus V_D\rvert\ge\theta$. 
	Since $|E_S|=|U_S|\times|V_S|-|\overline E_S|=\lvert U\setminus U_D\rvert\times\lvert V\setminus V_D\rvert-\lvert \overline E_{(U\cup V)\setminus D}\rvert$, and the complement vertex set of a maximum $k$-biplex of $G$ is always a minimal $k$-BDD of $\overline G$ according to Lemma \ref{lem:dual}, maximizing $|E_S|$ among all $k$-biplexes in $G$ is equivalent to maximizing $\lvert U\setminus U_D\rvert\times\lvert V\setminus V_D\rvert-\lvert \overline E_{(U\cup V)\setminus D}\rvert$ among all minimal $k$-BDDs in $\overline G$.
\end{proof}

Theorem~\ref{thm:dual} reformulates the maximum $k$-biplex search problem as finding a specific minimal $k$-BDD in the complement graph. Based on this formulation, we develop deletion-based branching and reduction strategies tailored to the maximum $k$-biplex search problem. Specifically, given a bipartite graph $G=(U,V,E)$ and its complement graph $\overline G=(U,V,\overline E)$, let $(D,S,C)$ be an instance of our deletion-based algorithm, where $D\subseteq U\cup V$ is the partial deletion set containing vertices that have been deleted from $\overline G$, $S\subseteq U\cup V$ is the partial $k$-biplex set containing vertices that cannot be deleted from $\overline G$ (i.e., the vertices that must be kept in the maximum $k$-biplex), and $C\subseteq U\cup V$ is the candidate set containing vertices to be moved to $D$ or $S$. The three sets are disjoint, and $G[S]$ is maintained as a valid $k$-biplex throughout the search. The instance $(D,S,C)$ aims to find a minimal $k$-BDD $D^*\subseteq C$ in $\overline G[S\cup C]$ such that $G[(U\cup V)\setminus (D\cup D^*)]$ is a maximum $k$-biplex of $G[S\cup C]$ containing $S$. In particular, the initial instance $((\emptyset,\emptyset),(\emptyset,\emptyset),(U,V))$ aims to find such a $k$-BDD in $\overline G$, which corresponds to a maximum $k$-biplex of $G$.

Given an instance $(D,S,C)$, our algorithm selects a vertex $u\in S\cup C$ and generate subbranches in the following way: if the selected vertex is in $C$, generate one subbranch that moves it to $D$; the remaining branches handle the cases where it is kept in $S$. For convenience, we refer to such a vertex $u$ as a branching vertex of the current instance $(D,S,C)$. Our algorithm then recursively applies this branching method until $D$ forms a minimal $k$-BDD of $\overline G$, where $G[S\cup C]$ gives a candidate solution for the maximum $k$-biplex of $G$. More specifically, our deletion-based branching strategy relies on the following branching rule.

\begin{theorem}[Deletion-based branching rule] \label{thm:delrule}
	Given a bipartite graph $G$ and a vertex $v$ of $G$, any $k$-BDD of $\overline G$ must contain either $v$ or at least $\overline d(v)-k$ of $v$'s neighbors in $\overline G$.
\end{theorem}

\begin{proof}
	Consider a $k$-BDD $D$ of $\overline G$. If $v\in D$, the theorem holds directly. Otherwise, $v$ remains in $\overline G\setminus D$. According to the definition of $k$-BDD, at most $k$ neighbors of $v$ in $\overline G$ can remain together with $v$. Therefore, at least $\overline d(v)-k$ neighbors of $v$ in $\overline G$ must be contained in $D$.
\end{proof}

According to Theorem~\ref{thm:delrule}, selecting a branching vertex $u\in C$ with a large degree in $\overline G[S\cup C]$ creates a strong branching condition: once $u$ is kept, at least $\overline d_{S\cup C}(u)-k$ vertices in $\overline N_C(u)$ must be deleted. Therefore, to delete as many vertices as possible from $C$, we can choose $u\in C$ with the maximum value of $\overline d_{S\cup C}(u)$. With this observation, we propose our deletion-based branching strategy as follows.

\noindent\textbf{Our deletion-based branching strategy.} Given the instance $(D,S,C)$, let $u=\mathrm{\arg\max}_{v\in S\cup C}\overline d_{S\cup C}(v)$ be the branching vertex. If $\overline d_{S\cup C}(u)>k$, let $r=\overline d_{S\cup C}(u)-k$ and $\overline N_C(u)=\{v_1,v_2,\dots,v_{\overline d_C(u)}\}$ in decreasing order of $\overline d_{S\cup C}(\cdot)$. The new subinstances are generated as follows.
\begin{enumerate}
	\item If $r\le 0$, $D$ forms a minimal $k$-BDD of $\overline G$, update the current best solution with $G[S\cup C]$ and return.
	\item If $u\in C$, generate one subinstance $(D\cup\{u\},S,C\setminus\{u\})$, corresponding to the case where $u$ is deleted. After this subinstance returns, add $u$ to $S$.
	\item For each $i=r,r+1,\ldots,\overline d_C(u)$, regard $v_i$ as the $r$-th deleted vertex in the order of $\overline N_C(u)$.  For every vertex subset $P\subseteq\{v_1,v_2,\dots,v_{i-1}\}$ with $|P|=r-1$, generate the subinstance $(D\cup P\cup\{v_i\},S\cup(\{v_1,v_2,\dots,v_{i-1}\}\setminus P),C\setminus\{v_1,v_2,\dots,v_i\})$, corresponding to the case where $v_i$ and $r-1$ vertices from $v_1,v_2,\dots,v_{i-1}$ are deleted.
\end{enumerate}

\begin{figure}[t]
	\centering
	\includegraphics[width=\columnwidth]{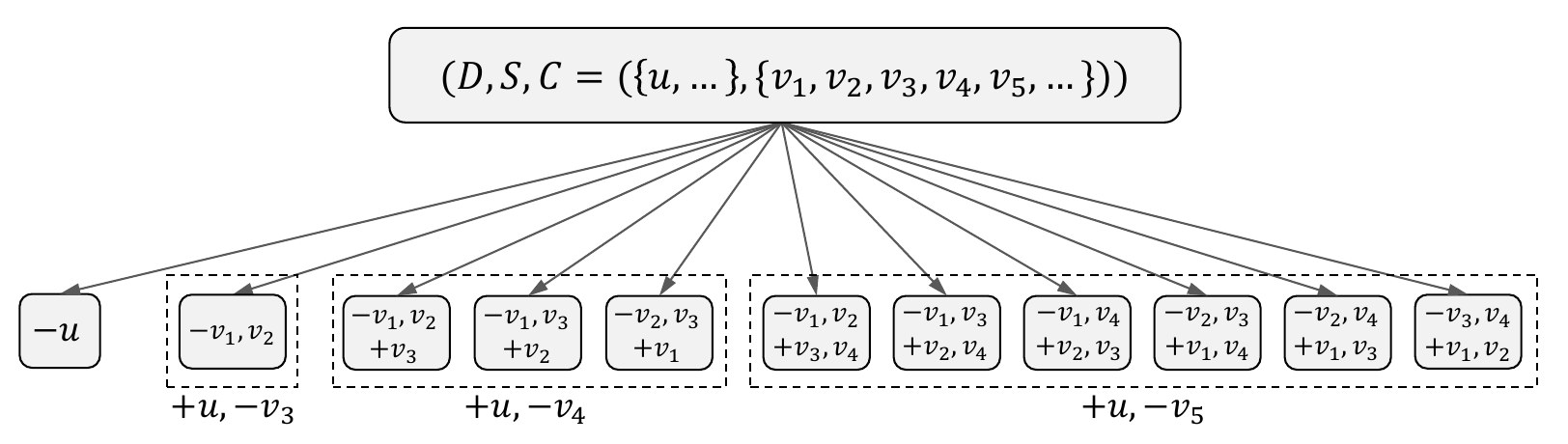}
	\caption{An illustration of our deletion-based branching strategy when $k=3$, $\overline d_{S\cup C}(u)=5$, and $\overline N_C(u)=\{v_1,v_2,v_3,v_4,v_5\}$. Here $+v$ means moving $v$ to $S$, and $-v$ means moving $v$ to $D$.}
	\label{fig:branching-example}
\end{figure}

Figure~\ref{fig:branching-example} illustrates the one-level branches generated by our deletion-based branching strategy. The first branch deletes the branching vertex $u$, while the remaining branches keep $u$ and delete two vertices from $\overline N_C(u)$. These branches are grouped by the last deleted vertex in the ordered set $\overline N_C(u)$.

\subsection{Branch Reduction Techniques} 

We introduce several branch reductions to further enhance the efficiency of our deletion-based branching strategy. First, we reduce $C$ through vertices with a large number of non-neighbors in $S$. 

\begin{lemma}[Branch Reduction 1] \label{lem:br1}
	Given an instance $(D,S,C)$ and a vertex $u\in C$, if $\overline d_S(u)>k$ or there exists $v\in \overline N_S(u)$ satisfying $\overline d_S(v)\ge k$, we can move $u$ from $C$ to $D$ directly.
\end{lemma}

\begin{proof}
	We prove it by contradiction. If $\overline d_S(u)>k$ and $u$ is not included in any $k$-BDD $D'$ derived from $(D,S,C)$, then $u$ will have more than $k$ neighbors in $\overline G[(S\cup C)\setminus D']$ and thus $D'$ is not a $k$-BDD of $\overline G$. Similarly, if there exists $v\in \overline N_S(u)$ with $\overline d_S(v)\ge k$, then $v$ will have at least $k+1$ neighbors in $\overline G[(S\cup C)\setminus D']$ and thereby $D'$ is not a $k$-BDD of $\overline G$ as well.
\end{proof}

%The above rule checks whether adding a candidate vertex would immediately violate the non-degree constraint. The following rule further uses the size constraint. If a candidate vertex does not have enough neighbors in the remaining graph, it cannot be contained in any $k$-biplex satisfying the size threshold $\theta$.

Then, we reduce $C$ through vertices that cannot be contained in any $k$-biplex satisfying the size threshold $\theta$.

\begin{lemma}[Branch Reduction 2] \label{lem:br2}
	Given an instance $(D,S,C)$ and a vertex $u\in C$, if $d_{S\cup C}(u)<\theta-k$, we can move $u$ from $C$ to $D$ directly.
\end{lemma}

\begin{proof}
	Assume that a $k$-biplex $S'$ derived from $(D,S,C)$ contains $u$. Since the opposite side of $u$ in $S'$ has at least $\theta$ vertices, vertex $u$ must have at least $\theta-k$ neighbors in $S'$. However, $S'\subseteq S\cup C$, and hence $d_{S'}(u)\le d_{S\cup C}(u)<\theta-k$, a contradiction. Therefore, $u$ cannot be contained in any solution and can be moved to $D$.
\end{proof}

Intuitively, if adding a candidate vertex from $C$ to $S$ does not make itself or any of its non-neighbors violate the $k$-biplex constraint, then keeping this vertex can only enlarge the solution. This leads to the following reduction.

\begin{lemma}[Branch Reduction 3] \label{lem:br3}
	Given an instance $(D,S,C)$ and a vertex $u\in C$, if $\forall v\in \{u\}\cup \overline N_{S\cup C}(u), \overline d_{S\cup C}(v)\le k$, we can move $u$ from $C$ to $S$ directly.
\end{lemma}

\begin{proof}
	Let $S'$ be any $k$-biplex derived from $(D,S,C)$ that does not contain $u$. We show that $S'\cup\{u\}$ is still a $k$-biplex. Since $\overline d_{S\cup C}(u)\le k$ and $S'\cup\{u\}\subseteq S\cup C$, vertex $u$ has at most $k$ non-neighbors in $S'\cup\{u\}$. Furthermore, for any vertex $v\in \overline N_{S'}(u)$, we have $\overline d_{S'\cup\{u\}}(v)\le \overline d_{S\cup C}(v)\le k$. The non-degrees of all other vertices in $S'$ remain unchanged after adding $u$ to $S'$. Thus, after adding $u$, $S'\cup\{u\}$ is still a $k$-biplex.
\end{proof}

We notice that the branching size can be further reduced for the case $\overline\delta_{S\cup C}=k+1$. In this case, every vertex in $\overline G[S\cup C]$ has at most $k+1$ neighbors. This branch reduction is based on the following lemma.

\begin{lemma}[Branch reduction 4] \label{lem:br4}
	Given an instance $(D,S,C)$ and a vertex $u\in C$. If $\overline d_{S\cup C}(u)\le k$, and $u$ has exactly one neighbor $v$ in $\overline G[S\cup C]$ satisfying $\overline d_{S\cup C}(v)>k$, and $v$ has exactly $k+1$ neighbors in $\overline G[S\cup C]$, then after moving $u$ to $D$, we can add $v$ and all vertices in $\overline N_{C}(v)\setminus\{u\}$ to $S$ directly.
\end{lemma}

\begin{proof}
	We prove it by contradiction. Let $D'$ be a minimal $k$-BDD derived from the current instance that contains $u$ and at least one vertex from $\{v\}\cup \overline N_C(v)\setminus\{u\}$. Let $S'=S\cup C\setminus D'$ be the $k$-biplex derived from deleting $D'$. If $v\in D'$, we have $\overline d_{S'}(u)< k$ and every neighbor $x$ of $u$ in $\overline G[S']$ satisfies $\overline d_{S'}(x)< k$. Therefore, $D'\setminus\{u\}$ is still a $k$-BDD and thus $D'$ is not minimal. If there exists a vertex $w\in \overline N_C(v)\setminus\{u\}$ and $w\in D'$, then we have $\overline d_{S'}(v)<k$, and every neighbor $x$ of $u$ in $\overline G[S']$ except $v$ satisfies $\overline d_{S'}(x)<k$. Therefore, $D'\setminus\{u\}$ is also a $k$-BDD and thus $D'$ is not minimal. 
\end{proof}

\begin{figure}
	\centering
	\includegraphics[width=\columnwidth]{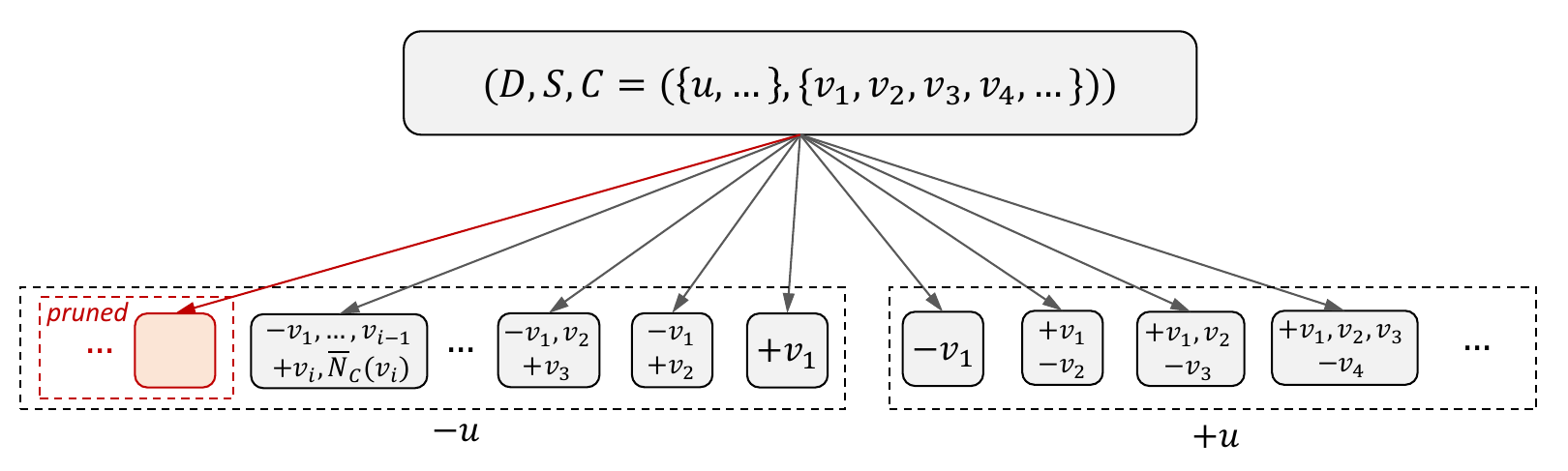}
	\caption{An illustration of our improved branching strategy for $\overline \delta_{S\cup C}=k+1$. Here $+v$ means moving $v$ to $S$, and $-v$ means moving $v$ to $D$. Red dashed boxes represent branches pruned by our improved branching strategy.}
	\label{fig:improved-branching-example}
\end{figure}

\noindent\textbf{Improved branching strategy for $\overline \delta_{S\cup C}=k+1$.} Based on Lemma~\ref{lem:br4}, we improve our deletion-based branching strategy for the case when $\overline \delta_{S\cup C}=k+1$ and $\max_{v\in S}\overline d_{S\cup C}(v)\le k$. In this case, the branching vertex $u$ is selected from $C$. We redesign the branching strategy when $u$ is deleted from $C$ to trigger Lemma~\ref{lem:br4} as early as possible. Specifically, let $u$ be the vertex in $\{v\in C\ \vert\ \overline d_{S\cup C}(v)=k+1\}$ with the minimum $\overline d_C(\cdot)$. Let $v_1,v_2,\dots,v_{\overline d_C(u)}$ be the vertices of $\overline N_C(u)$ in non-increasing order of $\overline d_{S\cup C}(\cdot)$. The subinstances are generated in the following way: for each $i$ from $1$ to $\overline d_C(u)$, we generate a subinstance:
\begin{equation*}
(D\cup\{u,v_1,v_2\dots,v_{i-1}\}, S\cup\{v_i\}, C\setminus\{u,v_1,v_2\dots,v_i\})
\end{equation*} 
until (1) $\overline d_{S\cup C\setminus\{v_1,v_2\dots,v_{i-1}\}}(u)\le k$, and (2) at most one vertex $v\in \{v_i,v_{i+1},\dots,v_{\overline d_{C}(u)}\}$ satisfies $\overline d_{S\cup C}(v)=k+1$. If $v_i$ is the only vertex in this set satisfying $\overline d_{S\cup C}(v_i)=k+1$, we generate one more subinstance according to Lemma \ref{lem:br4}: 
\begin{gather*}
	(D\cup\{u,v_1,v_2\dots,v_{i-1}\}, S\cup\{v_i\}\cup (\overline N_C(v_i)\setminus\{u\}),\\ C\setminus\{u,v_1,v_2\dots,v_i\}\setminus \overline N_C(v_i)).
\end{gather*} Otherwise, we terminate generating any new branches directly. These early termination can be established as the remaining subgraph $\overline G[S\cup C\setminus\{u,v_1,v_2,\dots,v_{i}\}]$ will not yield any minimal $k$-BDD according to Lemma \ref{lem:br4}.

Figure~\ref{fig:improved-branching-example} illustrates the one-level branches generated by our improved branching strategy. The left dashed box shows the branches where $u$ is moved to $D$, which are generated by this strategy. In these branches, the vertices of $\overline N_C(u)$ are considered in the given order, and each generated branch moves one vertex $v_i$ to $S$ while moving the preceding vertices $v_1,\ldots,v_{i-1}$ to $D$. Once Lemma~\ref{lem:br4} can be applied, i.e. $\overline d_{S\cup C}(u)\le k$ and $v_i$ is the last vertex in $\overline N_C(u)$ satisfying $\overline d_{S\cup C}(v_i)=k+1$, the remaining branches are pruned, and $v_i$ as well as all vertices in $\overline N_C(v_i)\setminus \{u\}$ are directly added to $S$. The right dashed box shows the branches where $u$ is added to $S$, which are generated by our deletion-based branching strategy. Since $\overline d_{S\cup C}(u)-k=1$, each such branch deletes exactly one vertex from $\overline N_C(u)$.

\subsection{Implementation details}

\begin{algorithm}[!t]
	\caption{The deletion-based algorithm}
	\label{alg:bdd}
	\smaller[1]
	\KwIn{Bipartite graph $G=(U,V,E)$, positive integers $k$ and $\theta$}
	\KwOut{A maximum $k$-biplex $G[S^*]$ s.t. $\lvert U_{S^*}\rvert\ge \theta$ and $\lvert V_{S^*}\rvert\ge \theta$}
	
	$S^*\leftarrow (\emptyset, \emptyset)$\;
	\bdd{$(\emptyset,\emptyset)$, $(\emptyset, \emptyset)$, $(U, V)$}\;
	
	\Fn{\bdd{$D$, $S$, $C$}}{
		
		Reduce $C$ based on Lemma \ref{lem:br1}, \ref{lem:br2} and \ref{lem:br3}\;
		%\If{\ub{$\overline G,D,S,C,k,\theta$}$\le |E_{S^*}|$}{\Return{}\;}
		
		\If{$\forall v\in S\cup C, \overline d_{S\cup C}(v)\le k$}{
			\lIf{$\lvert U_{D}\rvert\le \lvert U\rvert- \theta$ \And $\lvert V_{D}\rvert\le\lvert V\rvert-\theta$ \And $\lvert E_{S\cup C}\rvert>\lvert E_{S^*}\rvert$}{$S^*\leftarrow S\cup C$}
			%Update $S^*$ with $S\cup C$\;
			\Return{};	
		}
		
		\If{$\exists v\in S$ s.t. $\overline d_{S\cup C}(v)>k$}{$u\leftarrow\mathrm{\arg\max}_{v\in S}\overline d_{S\cup C}(v)$\;}
		\Else{$C_1\leftarrow\{v\in C\mid \overline d_{S\cup C}(v)=\max_{w\in C}\overline d_{S\cup C}(w)\}$\; $u\leftarrow\mathrm{\arg\min}_{v\in C_1}\overline d_C(v)$\;}
		%$\overline d\leftarrow \overline d_{S\cup C}(u)$\;
		%$\ell\leftarrow $\;
		
		$q\leftarrow\overline d_C(u)$\;
		$v_1,v_2,\dots,v_q\leftarrow\overline N_C(u)$ in decreasing order of $\overline d_{S\cup C}(\cdot)$\;
		
		\If{$u\in C$}{
			\lIf{$\overline d_{S\cup C}(u)=k+1$}{\bkpo{$D,S,C,u,(v_1,\dots,v_q)$}}
			\lElse{\bdd{$D\cup\{u\}$, $S$, $C\setminus\{u\}$}}
			$S\leftarrow S\cup\{u\}$; $C\leftarrow C\setminus\{u\}$\;
		}
		
		% \If{$\overline d-k>\overline d_C(u)$}{\Return{}\;}
		
		\For{$i\leftarrow\overline d_{S\cup C}(u)-k$ \To $q$}{
			\For{$P\subseteq\{v_1,v_2,\dots,v_{i-1}\}$  s.t. $\lvert P\rvert=\overline d_{S\cup C}(u)-k-1$}{
				$D'\leftarrow P\cup\{v_i\}$\;
				$S'\leftarrow \{v_1,v_2,\dots,v_i\}\setminus D'$\;
				\bdd{$D\cup D'$, $S\cup S'$, $C\setminus \{v_1,v_2,\dots,v_i\}$}\;
			}
		}
		
	}
	
	\Fn{\bkpo{$D,S,C,u,(v_1,\dots,v_q)$}}{
		\For{$i\leftarrow 1$ \To $q$}{
			$P_i\leftarrow\{v_1,v_2,\dots,v_{i-1}\}$;\quad $D_i\leftarrow\{u\}\cup P_i$\;
			\If{$\overline d_{S\cup(C\setminus P_i)}(u)\le k$ \And $(i=q$ \Or $\overline d_{S\cup C}(v_{i+1})\le k)$}{
				\If{$\overline d_{S\cup C}(v_i)=k+1$}{
					$S_i\leftarrow\{v_i\}\cup(\overline N_C(v_i)\setminus D_i)$\;
					\bdd{$D\cup D_i$, $S\cup S_i$, $C\setminus(D_i\cup S_i)$}\;
				}
				\Return{}\;
			}
			\bdd{$D\cup D_i$, $S\cup\{v_i\}$, $C\setminus(D_i\cup\{v_i\})$}\;
		}
	}
	
\end{algorithm}

%\begin{detail}
	Equipped with our deletion-based branching strategy and branch reduction techniques, we present the pseudocode of our deletion-based algorithm in Algorithm~\ref{alg:bdd}. It first initializes the best solution $S^*$ as an empty graph and invokes \bdd with all vertices in the candidate set (lines 1-2). For each instance $(D,S,C)$, the procedure \bdd exhaustively applies Lemma~\ref{lem:br1}, Lemma~\ref{lem:br2}, and Lemma~\ref{lem:br3} to reduce the candidate set (line 4). If all vertices in $S\cup C$ have at most $k$ non-neighbors in $S\cup C$, then $G[S\cup C]$ is already a valid $k$-biplex. In this case, the algorithm only needs to check the size constraint and update $S^*$ if more edges are obtained (lines 5-7).
	Otherwise, the algorithm selects a branching vertex $u$ and sorts the vertices in $\overline N_C(u)$ as $v_1,v_2,\dots,v_q$ in decreasing order of their non-degrees (lines 8-14). If some vertex in $S$ has non-degree larger than $k$, such a vertex with the maximum non-degree is selected; otherwise, $u$ is selected from $C$ with the maximum $\overline d_{S\cup C}(\cdot)$ and the minimum $\overline d_C(\cdot)$. If $u\in C$ and $\overline d_{S\cup C}(u)=k+1$, the algorithm invokes \bkpo to handle the branches where $u$ is moved to $D$ (lines 15-16). Specifically, \bkpo scans the ordered vertices and lets $P_i=\{v_1,\dots,v_{i-1}\}$ and $D_i=\{u\}\cup P_i$ (lines 25-26). Before the stopping condition is met, it generates the subinstance that moves $D_i$ to $D$ and fixes $v_i$ into $S$ (line 32). Once the remaining non-neighbors of $u$ satisfy the condition of Lemma~\ref{lem:br4}, \bkpo directly adds the corresponding vertices to $S$ and stops generating further branches (lines 27-31). After the branches with $u$ deleted are generated, the main procedure moves $u$ to $S$ (line 18) and applies the ordinary deletion-based branching rule (lines 19-23). Let $r=\overline d_{S\cup C}(u)-k$. For each possible position $i$ of the $r$-th deleted vertex in $\overline N_C(u)$, the algorithm chooses the other $r-1$ deleted vertices from $\{v_1,\dots,v_{i-1}\}$, deletes them together with $v_i$, and fixes the remaining prefix vertices into $S$ (lines 19-23).
%\end{detail}

We next analyze the worst-case time complexity of Algorithm~\ref{alg:bdd}.

\begin{theorem}
	Given a bipartite graph $G$ and two positive integers $k$ and $\theta$, the worst-case time complexity of Algorithm \ref{alg:bdd} is $O^*(\gamma_k^n)$, where $n=\lvert U\rvert+\lvert V\rvert$ and $\gamma_k<2$. In particular,  $\gamma_1=1.725$, $\gamma_2=1.856$, and $\gamma_3=1.928$.
\end{theorem}

\begin{proof}
	Let $T(n)$ denote the maximum number of leaf instances generated by Algorithm~\ref{alg:bdd}, where $n=|C|$ is the number of candidate vertices. In each recursive call, the procedure \bdd takes at most $O(m)$ time by maintaining degree values, selecting the branching vertex, and bucket-sorting the non-neighbors of the branching vertex. Therefore, the total running time of Algorithm~\ref{alg:bdd} is $O(m\cdot T(n))$.
	
	We next analyze $T(n)$ according to the branching strategy. Consider an instance after applying the reduction rules. Let $u$ be the selected branching vertex. Let $\overline d=\overline d_{S\cup C}(u)$, $q=\overline d_C(u)$ and $r=\overline d-k$. There are three possible cases to discuss.
	\begin{enumerate}
		\item \textbf{Case 1: $\overline d=k+1$ and $u\in C$.} Let $p$ be the number of vertices in $\overline N_C(u)$ whose non-degrees in $S\cup C$ are $k+1$. Since the vertices in $\overline N_C(u)$ are sorted in decreasing order of their non-degrees, these vertices are $v_1,\ldots,v_p$. In the $i$-th branch generated by \bkpo, it moves $D_i=\{u,v_1,\ldots,v_{i-1}\}$ to $D$ and fixes $v_i$ into $S$, thereby removing $i+1$ vertices from $C$.
		If $p\le 1$, at most one subinstance is generated in \bkpo, which moves $u$ to $D$ and $v_1$ to $S$. At most $k+1$ subinstances are generated by \bdd, each of which moves $u$ to $S$ and moves one vertex from $\overline N_C(u)$ to $D$. Therefore, the recurrence is given by 
		\begin{equation}
			T(n)\le T(n-2) + \sum_{i=1}^{k+1} T(n-i-1).
		\end{equation}
		If $p\ge2$, then \bkpo generates subinstances for $i=1,\ldots,p-1$, and applies Lemma~\ref{lem:br4} at $v_p$. Since $u$ is chosen with the minimum $\overline d_C(\cdot)$ among vertices of non-degree $k+1$, we have $\overline d_C(v_p)\ge \overline d_C(u)=q$. Moreover, all vertices in $\overline N_C(u)$ are on the same side as $v_p$, and hence the only vertex in $D_p$ adjacent to $v_p$ in $\overline G$ is $u$. Thus, the branch generated by Lemma~\ref{lem:br4} removes at least $p+q$ vertices from $C$. After \bkpo finishes, the main procedure moves $u$ to $S$. Since $\overline d=k+1$, the ordinary deletion-based branching rule further generates $q$ branches, where the $i$-th branch removes $i+1$ vertices from $C$. Therefore, the recurrence is given by 
		\begin{align}
			T(n)\le&
			\sum_{i=1}^{q}T(n-i-1)
			+
			\sum_{i=2}^{p}T(n-i)
			+
			T(n-p-q).
			\label{eq:kplusone-c}
		\end{align}
		Among all $1\le p\le q\le k+1$, the largest branching factor of Eq.~(\ref{eq:kplusone-c}) is obtained when $q=p=k+1$. Hence, this case is bounded by
		\begin{align}
			T(n)\le&
			2\sum_{i=2}^{k+1}T(n-i)
			+
			T(n-k-2)
			+
			T(n-2k-2) \nonumber \\
			\le&
			\sum_{i=1}^{k}T(n-i)
			+
			T(n-k-2)
			+
			T(n-2k-2)
			\label{eq:kplusone-c-bound}
		\end{align}
		
		\item \textbf{Case 2: $\overline d=k+1$ and $u\in S$.} Since $u$ is already contained in $S$, the algorithm does not generate the branch that moves $u$ to $D$. The ordinary deletion-based branching rule only needs to delete one vertex from $\overline N_C(u)$. Thus, the recurrence is given by
		\begin{equation}
			T(n)\le \sum_{i=1}^{k+1}T(n-i).
			\label{eq:kplusone-s}
		\end{equation}
		
		Note that Eq.~(\ref{eq:kplusone-c-bound}) is dominated by Eq.~(\ref{eq:kplusone-s}). Let $\eta_k$ be the largest real root of $x^n=\sum_{i=1}^{k+1}x^{n-i}$.
		Then this case runs in $O^*(\eta_k^n)$ time, where $\eta_1=1.618$, $\eta_2=1.839$, and $\eta_3=1.928$.
		
		\item \textbf{Case 3: $\overline d\ge k+2$.} If $u\in C$, the branch that moves $u$ to $D$ contributes $T(n-1)$; if $u\in S$, this branch is absent and the recurrence is dominated by the former case. In the branches where $u$ is kept, the algorithm chooses the position $i$ of the $r$-th deleted vertex in $\overline N_C(u)$ and chooses the other $r-1$ deleted vertices from $\{v_1,\ldots,v_{i-1}\}$. Therefore,
		\begin{equation}
			T(n)\le T(n-1)+\sum_{i=r}^{q} \binom{i-1}{r-1} T(n-i-1).
			\label{eq:bdd-rec}
		\end{equation}
		For $\overline d\ge k+2$, the largest branching factor is obtained when $r=2$ and $q=k+2$. Thus, Eq.~(\ref{eq:bdd-rec}) is bounded by
		\begin{equation}
			T(n)\le T(n-1)+\sum_{i=2}^{k+2}(i-1)T(n-i-1).
			\label{eq:large-degree-bound}
		\end{equation}
		Let $\zeta_k$ be the largest real root of $x^{k+3}=x^{k+2}+\sum_{j=1}^{k+1} jx^{k+1-j}$.
		Then this case runs in $O^*(\zeta_k^n)$ time, where $\zeta_1=1.725$, $\zeta_2=1.856$, and $\zeta_3=1.923$.
	\end{enumerate}
	
	Combining the above cases, the running time is $O^*(\gamma_k^n)$, where $\gamma_k=\max\{\eta_k,\zeta_k\}$. Therefore, $\gamma_1=1.725$, $\gamma_2=1.856$, and $\gamma_3=1.928$. This completes the proof.
\end{proof}

\section{Optimization Techniques}

In this section, we propose several optimization techniques to further enhance the practical performance of our deletion-based algorithm, including efficient upper-bounding techniques to prune unnecessary instances in linear time and a new heuristic algorithm to identify an initial near-maximum $k$-biplex in polynomial time.

\subsection {Novel Upper-Bounding Techniques}
\label{sec:ub}

In this subsection, we propose an efficient upper-bounding technique to prune
instances that cannot produce a $k$-biplex with more edges than $S^*$. Given an
instance $(D,S,C)$, let $S=(U_S,V_S)$ and $C=(U_C,V_C)$. 
We first compute the degree-based vertex upper bounds for both sides as
$c_U=\min_{v\in V_S}(d_{S\cup C}(v)+k)$ and $c_V=\min_{v\in U_S}(d_{S\cup C}(v)+k)$.
If $c_U<\theta$ or $c_V<\theta$, the instance can be safely pruned.

Next, we derive an edge upper bound from the vertex newly added to $S$. Suppose that a vertex $u$ has just been moved from $U_C$ to $U_S$. The case that $u$ is moved from $V_C$ to $V_S$ is symmetric.
Since $u$ has been added to $U_S$, at most $k-\overline d_S(u)$ vertices in $\overline N_C(u)$ can be kept in any $k$-biplex derived from $(D,S,C)$. 
For a vertex $v\in V_S\cup V_C$, let $f(v)=\min\{d_{S\cup C}(v),c_U\}$. Clearly, $f(v)$ provides an upper bound for $v$'s degree in any derived $k$-biplex, i.e., adding $v$ to $S$ will introduce at most $f(v)$ edges to the final $k$-biplex. Therefore, an edge upper bound can be obtained by keeping all vertices in $V_C\setminus\overline N_C(u)$ and only $k-\overline d_S(u)$ vertices with the largest $f(\cdot)$ values in $\overline N_C(u)$, as stated in the following lemma.

\begin{lemma} [Vertex-based upper bound] \label{lem:ub1}
	Given an instance $(D,S,C)$ and a vertex $u$ that is newly added to $S$, let $v_1,v_2,\dots,v_{\overline d_{C}(u)}$ be the vertices of $\overline N_C(u)$ in non-increasing order with respect to $f(v)$. For any $k$-biplex $S'$ derived from $(D,S,C)$, we have that:
	\begin{equation} \label{eq:ub1}
	\lvert E_{S'}\rvert\le \sum_{v\in V_S\cup N_C(u)} f(v) + \sum_{i=1}^{k-\overline d_S(u)} f(v_i).
	\end{equation}
\end{lemma}

\begin{proof}
	
	For each vertex $v$ on the opposite side of $u$, its contribution to
	$\lvert E_{S'}\rvert$ is at most $f(v)$. As $u$ is added to $S$, at most $k-\overline d_S(u)$ non-neighbors of $u$ can still be added to $S$. Since $f(v)$ is independent of how vertices are chosen from $V_S\cup V_C$, vertices in $V_S\cup N_C(u)$ contribute at most $\sum_{v\in V_S\cup N_C(u)} f(v)$ edges to $S'$, and vertices in $\overline N_C(u)$ contribute at most $\sum_{i=1}^{k-\overline d_S(u)} f(v_i)$ edges to $S'$.
\end{proof}

\begin{figure}[!t]
	\centering
	\smaller[1]
	\begin{minipage}[b]{0.5\columnwidth}
		\centering
		\resizebox{\linewidth}{!}{
			\begin{tikzpicture}[vtx/.style={circle, draw, inner sep=0pt, minimum size=12pt, font=\small}, lab/.style={draw=none, font=\small}]
				\node[vtx] (u1) at (-2.10,1.65) {$\mathrm{u}_1$};
				\node[vtx, fill=red!12] (u2) at (-1.35,1.65) {$\mathrm{u}_2$};
				\node[vtx] (u3) at (-0.35,1.65) {$\mathrm{u}_3$};
				\node[vtx] (u4) at (0.45,1.65) {$\mathrm{u}_4$};
				\node[vtx] (u5) at (1.25,1.65) {$\mathrm{u}_5$};
				\node[vtx] (v1) at (-2.10,0) {$\mathrm{v}_1$};
				\node[vtx] (v2) at (-1.35,0) {$\mathrm{v}_2$};
				\node[vtx] (v3) at (-0.45,0) {$\mathrm{v}_3$};
				\node[vtx] (v4) at (0.25,0) {$\mathrm{v}_4$};
				\node[vtx] (v5) at (0.95,0) {$\mathrm{v}_5$};
				\node[vtx] (v6) at (1.65,0) {$\mathrm{v}_6$};
				\node[vtx] (v7) at (2.35,0) {$\mathrm{v}_7$};
				\draw[gray!65, dashed, line width=0.25pt] (-0.90,-0.50) -- (-0.90,1.95);
				\node[lab] at (-1.73,2.22) {$S$};
				\node[lab] at (0.95,2.22) {$C$};
				\foreach \a/\b in {u1/v3,u1/v4,u1/v5,u2/v1,u2/v2,u2/v3,u2/v4,u3/v1,u3/v2,u3/v3,u3/v4,u3/v5,u3/v6,u3/v7,u4/v1,u4/v2,u4/v3,u4/v4,u4/v5,u4/v6,u4/v7,u5/v1,u5/v3,u5/v6,u5/v7}
				\draw[gray!62, line width=0.25pt] (\a) -- (\b);
				\draw[red, dashed, line width=0.3pt, bend left=10] (u2) to (v5);
				\draw[red, dashed, line width=0.3pt, bend left=8] (u2) to (v6);
				\draw[red, dashed, line width=0.3pt, bend left=6] (u2) to (v7);
				\node[lab] at (-2.75,1.65) {$U$};
				\node[lab] at (-2.75,0) {$V$};
			\end{tikzpicture}
		}
		\vspace{5pt}
		\centerline{(a) The example graph}
	\end{minipage}
	\hfill
	\begin{minipage}[b]{0.49\columnwidth}
		\centering
		\setlength{\tabcolsep}{0.8pt}
		\renewcommand{\arraystretch}{0.98}
		\begin{tabular}{cc|cc|ccccc}
			& & \multicolumn{2}{c|}{$V_S$} & \multicolumn{5}{c}{$V_C$}\\
			& & $v_1$ & $v_2$ & $v_3$ & $v_4$ & $v_5$ & $v_6$ & $v_7$\\
			\hline
			\multirow{2}{*}{$U_S$} & $u_1$ & $\circ$ & $\circ$ & $\bullet$ & $\bullet$ & $\bullet$ & $\circ$ & $\circ$\\
			& $u_2$ & $\bullet$ & $\bullet$ & $\bullet$ & $\bullet$ & \textcolor{red}{$\circ$} & \textcolor{red}{$\circ$} & \textcolor{red}{$\circ$}\\
			\hline
			\multirow{3}{*}{$U_C$} & $u_3$ & $\bullet$ & $\bullet$ & $\bullet$ & $\bullet$ & $\bullet$ & $\bullet$ & $\bullet$\\
			& $u_4$ & $\bullet$ & $\bullet$ & $\bullet$ & $\bullet$ & $\bullet$ & $\bullet$ & $\bullet$\\
			& $u_5$ & $\bullet$ & $\circ$ & $\bullet$ & $\circ$ & $\circ$ & $\bullet$ & $\bullet$\\
			\hline
			\multicolumn{2}{c|}{$f(v)$} & $4$ & $3$ & $5$ & $4$ & $3$ & $3$ & $3$\\
		\end{tabular}
		\vspace{5pt}
		\centerline{(b) The adjacent matrix}
	\end{minipage}
	\caption{An example instance for upper-bounding techniques when $k=2$. The left part shows the bipartite graph, where dashed red lines denote the non-edges between the newly added vertex $u_2$ (red color) and $\overline N_C(u_2)$. In the right matrix, $\bullet$ denotes an edge in $G[S\cup C]$ and $\circ$ denotes a non-edge.}
	\label{fig:ub-example}
\end{figure}
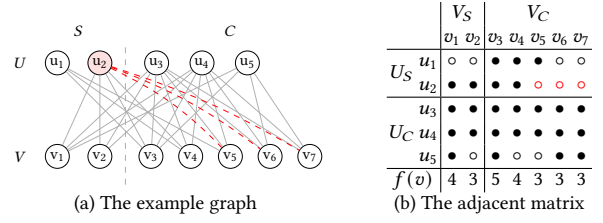

We notice that this upper bound may not be tight enough, as it only restricts $u$'s non-neighbors in $C$ without considering $u$'s neighbors in $C$. To address this, we further propose an edge upper bound from $U_S$. The case for $V_S$ is symmetric.
Specifically, for a vertex $v\in V_C$, adding $v$ to $S$ introduces
$\overline d_S(v)$ non-neighbors to vertices in $U_S$. Since every vertex in
$U_S$ can have at most $k$ non-neighbors, the total number of such introduced
non-neighbors is at most $\sum_{w\in U_S}(k-\overline d_S(w))$.
Therefore, if all vertices in $V_C$ introduce more non-neighbors than the above
allowance, some vertices in $V_C$ must be discarded.

\begin{lemma} [Set-based upper bound]
	\label{lem:ub2}
	
Given an instance $(D,S,C)$, let $x_1,x_2,\dots,x_{|V_C|}$ be the vertices of $V_C$ in non-increasing order
with respect to $\overline d_S(x)$. Let $p$ be the smallest integer satisfying
\begin{equation}
\sum_{i=1}^{p}\overline d_S(x_i)
\ge
\sum_{v\in V_C}\overline d_S(v)
-
\sum_{w\in U_S}(k-\overline d_S(w)).
\end{equation}
Let $y_1,y_2,\dots,y_{|V_C|}$ be the vertices of $V_C$ in non-increasing order with
respect to $f(v)$. We can derive the following upper bound.
For any $k$-biplex $S'$ derived from $(D,S,C)$, we have that:
	\begin{equation} \label{eq:ub2}
		\lvert E_{S'}\rvert
		\le
		\sum_{v\in V_S} f(v)
		+
		\sum_{i=1}^{\lvert V_C\rvert-p} f(y_i).
	\end{equation}
\end{lemma}

\begin{proof}
	For any $k$-biplex $S'$ derived from $(D,S,C)$, the total number of non-neighbors introduced by
	vertices selected from $V_C$ cannot exceed
	$\sum_{w\in U_S}(k-\overline d_S(w))$. According to the definition of $p$,
	at least $p$ vertices in $V_C$ cannot be selected. Therefore, at most
	$|V_C|-p$ vertices from $V_C$ can be contained in $S'$.
	For each vertex $v\in V_S\cup V_C$, its contribution to
	$\lvert E_{S'}\rvert$ is at most $f(v)$. Thus, vertices in $V_S$ contribute at most $\sum_{v\in V_S} f(v)$ edges to $S'$, and vertices in $V_C$ contribute at most $\sum_{i=1}^{\lvert V_C\rvert-p} f(y_i)$ edges to $S'$.
\end{proof}

\begin{example}
	Consider the instance shown in Fig.~\ref{fig:ub-example}, where $k=2$, $S=(U_S,V_S)$ with $U_S=\{u_1,u_2\}$ and $V_S=\{v_1,v_2\}$, and $C=(U_C,V_C)$ with $U_C=\{u_3,u_4,u_5\}$ and $V_C=\{v_3,v_4,v_5,v_6,v_7\}$. Assume that $u_2$ has just been moved from $U_C$ to $U_S$. We have $c_U=5$ and $c_V=5$, and the corresponding $f(\cdot)$ values of vertices in $V_S\cup V_C$ are shown in the last row of Fig.~\ref{fig:ub-example}.
	For the vertex-based upper bound, $\overline d_S(u_2)=0$ and $\overline N_C(u_2)=\{v_5,v_6,v_7\}$. Thus, any $2$-biplex containing $u_2$ can keep at most two vertices from $\overline N_C(u_2)$. By Lemma~\ref{lem:ub1}, we have
	\[
	e_1=(f(v_1)+f(v_2)+f(v_3)+f(v_4))+(f(v_5)+f(v_6))=22.
	\]
	For the set-based upper bound, the vertices $v_3,v_4,v_5,v_6,v_7$ introduce $0,0,1,2,2$ non-neighbors to $U_S$, respectively. The remaining number of non-neighbors allowed by $U_S$ is $(2-\overline d_S(u_1))+(2-\overline d_S(u_2))=0+2=2$. Hence, at least two vertices in $V_C$ cannot be selected. Keeping the three vertices in $V_C$ with the largest $f(\cdot)$ values gives
	\[
	e_2=f(v_1)+f(v_2)+f(v_3)+f(v_4)+f(v_5)=19.
	\]
	Therefore, Algorithm~\ref{alg:ub} returns $\min\{c_U\cdot c_V,e_1,e_2\}=19$. If the current best solution already contains at least $19$ edges, this instance can be safely pruned.
\end{example}

\begin{algorithm}[!t]
	\caption{\protect\ub{$D$, $S$, $C$, $u$}}
	\label{alg:ub}
	\smaller[0.5]
	\KwIn{Instance $(D,S,C)$ after moving $u$ from $U_C$ to $U_S$}
	\KwOut{An edge upper bound}
	
	%\If{$\exists v\in S$ s.t. $\overline d_S(v)>k$}{\Return{$-\infty$}\;}
	$c_U\leftarrow \min\{|U_S|+|U_C|,\min_{v\in V_S}(d_{S\cup C}(v)+k)\}$\;
	$c_V\leftarrow \min\{|V_S|+|V_C|,\min_{v\in U_S}(d_{S\cup C}(v)+k)\}$\;
	\lIf{$c_U<\theta$ \Or $c_V<\theta$}{\Return{$0$}}
	
	\lFor{$v\in U_S\cup U_C$}{$f(v)\leftarrow \min\{d_{S\cup C}(v), c_V\}$}
	\lFor{$v\in V_S\cup V_C$}{$f(v)\leftarrow \min\{d_{S\cup C}(v), c_U\}$}
	%\If{$b<0$}{\Return{$-\infty$}\;}
	
	$e_1\leftarrow 0$\;
	\lFor{$v\in V_S\cup N_C(u)$}{$e_1\leftarrow e_1+f(v)$}
	$v_1,v_2,\dots,v_{\overline d_C(u)}\leftarrow$ $\overline N_C(u)$ in non-increasing order of $f(\cdot)$\;
	\lFor{$i\leftarrow 1$ \To $\min\{k-\overline d_S(u), \overline d_C(u)\}$}{$e_1\leftarrow e_1+ f(v_i)$}
	
	$e_2\leftarrow\min\{$ \setub{$U_S$, $V_S$, $U_C$, $V_C$}, \setub{$V_S$, $U_S$, $V_C$, $U_C$} $\}$\;
	
	\Return{$\min\{c_U\cdot c_V$, $e_1$, $e_2\}$}\;
	
	\Fn{\setub{$U_S$,$V_S$,$U_C$,$V_C$}}{
	
	$r\leftarrow 0$;\ \ $e\leftarrow 0$;\ \ $p\leftarrow 0$\;
	\lFor{$v\in V_C$}{$r\leftarrow r+\overline d_S(v)$}
	\lFor{$w\in U_S$}{$r\leftarrow r-(k-\overline d_S(w))$}
	$x_1,x_2,\dots,x_{\lvert V_C\rvert}\leftarrow$ $V_C$ in non-increasing order of $\overline d_S(\cdot)$\;
	\While{$p<|V_C|$ \And $r>0$}{
		$p\leftarrow p+1$\;
		$r\leftarrow r-\overline d_S(x_p)$\;
	}
	\lFor{$v\in V_S$}{$e\leftarrow e+f(v)$}
	$y_1,y_2,\dots,y_{\lvert V_C\rvert}\leftarrow$ $V_C$ in non-increasing order of $f(\cdot)$\;
	\lFor{$i\leftarrow 1$ \To $\lvert V_C\rvert-p$}{$e\leftarrow e+f(y_i)$}
	
	\Return{$e$}\;
	}
\end{algorithm}

\begin{detail}
Algorithm~\ref{alg:ub} first computes the vertex upper bounds $c_U$ and $c_V$ (lines 1-2).
If either of them is smaller than $\theta$, the current instance can be pruned (line 3).
Then, it computes the degree upper bound $f(\cdot)$ for each vertex (lines 4-5).
After that, Algorithm~\ref{alg:ub} computes two edge upper bounds. It first computes the vertex-based upper bound according to Lemma~\ref{lem:ub1}. Specifically, it keeps all vertices in $V_S\cup N_C(u)$ (line 7), and selects at most
$\min\{k-\overline d_S(u),\overline d_C(u)\}$ vertices with the largest $f(\cdot)$ values from $\overline N_C(u)$ (lines 8-9).
Then it computes two set-based upper bounds, which are derived from calling
\setub twice, once for each side (line 10). In each call, it first computes
the excess number of non-neighbors of vertices in $U_S$ caused by adding all vertices from $V_C$ (lines 14-15).
Then it removes vertices from $V_C$ with the largest $\overline d_S(\cdot)$
values until the excess is covered (lines 16-19). Finally, it keeps the remaining number of vertices in $V_C$ with the largest $f(\cdot)$ and all vertices in $V_S$ to obtain the set-based upper bound (lines 20-22).
\end{detail}

\begin{lemma}
	Given an instance $(D,S,C)$ and a vertex $u$ that is newly added to $S$, Algorithm~\ref{alg:ub} runs in $O(|S|+|C|)$ time.
\end{lemma}

\begin{proof}
	In Algorithm~\ref{alg:ub}, the values $c_U$, $c_V$, and $f(\cdot)$ can be computed by traversing vertices in $S\cup C$, which takes $O(|S|+|C|)$ time. In computing $e_1$, the vertices in
	$\overline N_C(u)$ are sorted by bucket sort, and the remaining operations are
	linear traversals. Thus, it takes $O(|S|+|C|)$ time.
	In \setub, the values $r$ and $e$ are computed by linear traversals.
	The two orderings can be obtained by bucket sort, since both
	$\overline d_S(\cdot)$ and $f(\cdot)$ are integers bounded by $|S|+|C|$. The value of $p$ is also computed in at most $\lvert V_C\rvert$ times. Thus, \setub takes $O(|S|+|C|)$ time. Based on the above analysis, the total time complexity of
	Algorithm~\ref{alg:ub} is $O(|S|+|C|)$.
\end{proof}

\subsection{A New Heuristic Algorithm}

\begin{algorithm}[t]
	\caption{\protect\heu{$G$, $\theta$, $k$}}
	\label{alg:heu}
	\smaller[0.5]
	\KwIn{Bipartite graph $G=(U,V,E)$, positive integers $\theta$ and $k$}
	\KwOut{A large $k$-biplex $G[S^*]$ with $\lvert U_{S^*}\rvert\ge \theta$ and $\lvert V_{S^*}\rvert\ge \theta$}
	
	$S^*\leftarrow (U,V)$\;
	\While{$G[S^*]$ is not a $k$-biplex}{
		\lIf{$\lvert U_{S^*}\rvert>\lvert V_{S^*}\rvert$}{$u\leftarrow \mathrm{\arg\min}_{v\in U_{S^*}} d_{S^*}(v)$}
		\lElse{$u\leftarrow \mathrm{\arg\min}_{v\in V_{S^*}} d_{S^*}(v)$}
		$S^*\leftarrow S^*\setminus\{u\}$\;
	}
	
	\lIf{$\lvert U_{S^*}\rvert<\theta$ \Or $\lvert V_{S^*}\rvert<\theta$}{$S^*\leftarrow(\emptyset,\emptyset)$}

	%$\theta_U\leftarrow \max_{v\in V}d(v)+k$\;
	%\While{$\theta_U>\theta$}{
		%$\theta_V\leftarrow \max\{\theta, \lfloor\lvert E_S\rvert/\theta_U\rfloor\}$; $\theta_U\leftarrow \max\{\theta, \lfloor\theta_U/2\rfloor\}$\;
		
		%$G'=(U',V',E')\leftarrow$ the $(\theta_V-k, \theta_U-k)$-core of $G$\;
		$u_1,u_2,\dots,u_{\lvert U\rvert}\leftarrow$ $U$ in non-decreasing degree order\; 
		\For{$i\leftarrow 1\dots \lvert U\rvert$}{
			$S\leftarrow (\{u_i\},\emptyset)$; $C\leftarrow (N^{2+}(u_i), N(u_i)\cup N^{3}(u_i))$\;
			\While{$C\not=\emptyset$}{
				Reduce $C$ with Lemma \ref{lem:br1}, \ref{lem:br2} and \ref{lem:br3}; break if $C=\emptyset$\;
				$u\leftarrow\mathrm{\arg\max}_{v\in C} d_{S\cup C}(v)$\;
				\If{$G[S\cup\{u\}]$ is a $k$-biplex}{$S\leftarrow S\cup\{u\}$\;}
				$C\leftarrow C\setminus\{u\}$\;
			}
			\lIf{$\lvert U_S\rvert\ge\theta$ \And $\lvert V_S\rvert\ge\theta$ \And $\lvert E_S\rvert>\lvert E_{S^*}\rvert$}{$S^*\leftarrow S$}
		}
		%}
	
	\Return{$G[S^*]$}\;
	
\end{algorithm}

We propose a heuristic algorithm to efficiently identify an initial large $k$-biplex, which provides a tight lower bound on the edge size of the maximum $k$-biplex and helps prune the non-maximum instances. The pseudocode of this algorithm is shown in Algorithm~\ref{alg:heu}. Specifically, the algorithm consists of two stages. The first stage (lines 1-6) aims to extract a valid $k$-biplex by a greedy deletion strategy. Starting with the entire graph $G$, the algorithm iteratively removes the vertex with the minimum degree. To prevent the resulting structure from becoming heavily skewed, this deletion is strictly restricted to the partition currently containing more vertices (lines 3-4). The process terminates when the remaining subgraph $G[S^*]$ forms a valid $k$-biplex (line 2). If $G[S^*]$ violates the size threshold $\theta$, $S^*$ is set to empty (line 6).

However, this greedy strategy may prematurely delete useful vertices and thereby miss a large $k$-biplex in practice. To address this problem, we propose the second stage (lines 7-15) to systematically explore localized subgraphs. Specifically, the algorithm first sorts vertices of $U$ as $u_1,u_2,\dots,u_{|U|}$ in ascending degree order (line 7). For each vertex $u_i\in U$, the algorithm greedily constructs a valid $k$-biplex that contains $u_i$ while excluding $u_1,u_2,\dots,u_{i-1}$. We first present the following lemma to constrain the distance between two vertices of a $k$-biplex.

\begin{lemma} \cite{RN66} \label{lem:dis}
	The distance between any two vertices of a $k$-biplex $S=(U_S,V_S)$ is at most $3$ if $|U_S|\ge 2k+1$ and $|V_S|\ge 2k+1$.
\end{lemma}

Let $N^2(u_i)=\{w\in N(v)\mid v\in N(u_i)\}$ and $N^3(u_i)=\{x \in N(w)\mid w\in N^2(u_i)\}$ denote the set of $u_i$'s 2-hop and 3-hop neighbors, respectively. Let $N^{2+}(u_i)=N^2(u_i)\cap\{u_{i+1},u_{i+2},\dots,u_{|U|}\}$. According to Lemma~\ref{lem:dis}, any $k$-biplex whose first vertex in the above ordering is $u_i$ must be contained in $\{u_i\}\cup N^{2+}(u_i)\cup N(u_i)\cup N^3(u_i)$. Therefore, the algorithm initializes $S$ as $(\{u_i\},\emptyset)$ and  $C$ as the vertices within distance at most $3$ from $u_i$ after removing lower-order vertices (line 9). Then, the algorithm repeatedly reduces $C$; if $C$ becomes empty, the current expansion is terminated. Otherwise, the vertex in $C$ with the largest degree in $G[S\cup C]$ is selected, moved to $S$ only when the resulting subgraph remains a $k$-biplex, and then discarded from $C$ (lines 10-14). Finally, the algorithm updates $S^*$ if the obtained $G[S]$ has more edges than $G[S^*]$ (line 15).

\begin{theorem}
	Given a bipartite graph $G$ and two integers $\theta$ and $k$, Algorithm \ref{alg:heu} runs in $O(mn)$ time.
\end{theorem}

\begin{proof}
	In Algorithm \ref{alg:heu}, the first stage (lines 2-5) can be implemented using a peeling method adapted from the core decomposition \cite{RN78}, which takes $O(m)$ time. In the second stage, bucket-sorting vertices of $U$ by degree takes $O(n)$ time (line 7). For each vertex $u_i$, initializing $S$ and $C$ (line 9) takes $O(n\delta)=O(m)$ time. During the greedy expansion (lines 10-14), the algorithm maintains the degree value $d_{S\cup C}(\cdot)$ for vertices in $C$. When a vertex is removed from $C$, updating the degree value takes $O(\delta)$ time. By using a peeling method similar to the first stage, selecting a vertex with the largest $d_{S\cup C}(\cdot)$ takes only $O(1)$ time (line 12). Therefore, the greedy expansion loop takes totally $O(n\delta)=O(m)$ time.
	As $|U|\le n$, the total time of the second stage is $O(nm)$. Consequently, the total running time of Algorithm~\ref{alg:heu} is $O(m+mn)=O(mn)$.
\end{proof}

\begin{table}[!t]
	\centering
	\caption{Datasets used in experiments}
	\label{tab:datasets}
	\resizebox{0.48\textwidth}{!}{
		\begin{tabular}{lcccc}
			\toprule
			Dataset & $\lvert U\rvert$ & $\lvert V\rvert$ & $\lvert E\rvert$ & $\delta$ \\
			\midrule
			Youtube		&94,238		&30,087		&293,360	&(1,035, 7,591)					\\
			LKML		&42,045		&337,509	&599,858	&(31,719, 6,627)	\\
			%Team		&901,130	&34,461		&1,366,466	&(17, 2,671)		\\
			Mummun		&175,214	&530,418	&1,890,661	&(968, 19,805)		\\
			Citeu		&153,277	&731,769	&2,338,554	&(189,292, 1,264)	\\
			IMDB		&303,617	&896,302	&3,782,463	&(1,334, 1,590)		\\
			Amazon		&2,146,057	&1,230,915	&5,743,258	&(12,180, 3,096)	\\
			%Twitter		&244,537	&9,129,669	&10,214,177	&(640, 18,874)		\\
			Aol			&4,811,647	&1,632,788	&10,741,953	&(100,629, 84,530)	\\
			%DBLP		&1,953,085	&5,624,219	&12,282,059	&(1,386, 287)					\\
			Google		&5,998,790	&4,443,631	&20,592,962	&(423, 95,165)		\\
			\bottomrule
		\end{tabular}
	}
\end{table}

\section{Experiments}

\begin{table*}[!t]
	\centering
	\small
	\caption{Efficiency comparison among different algorithms on real-world graphs (in seconds)}
	\label{tab:overall-runtime}
	\setlength{\tabcolsep}{4pt}
	\resizebox{\textwidth}{!}{\begin{tabular}{c|c|c|c|ccc|c|c|c|ccc|c|c|c|ccc}
			%\toprule
			\hline
			%\cline{1-22}
			
			Dataset&$k$&$\theta$&$\lvert E^*\rvert$&\textsf{DMBP}&\fastbb&\cpc&$k$&$\theta$&$\lvert E^*\rvert$&\textsf{DMBP}&\fastbb&\cpc&$k$&$\theta$&$\lvert E^*\rvert$&\textsf{DMBP}&\fastbb&\cpc\\
			%\cline{1-22}
			\hline
			\multirow{4}{*}{Youtube} 
			& \multirow{2}{*}{1} & 3 & 945 & \textbf{35} & 1,232 & 5,510 & \multirow{2}{*}{2} & 7 & 280 & \textbf{5,500} & - & - & \multirow{2}{*}{3} & 10 & 345 & \textbf{9,452} & - & - \\
			&  & 6 & 302 & \textbf{7.66} & 186 & 162 &  & 10 & 275 & \textbf{186} & 16,675 & 6,951 &  & 13 & 345 & \textbf{347} & - & - \\
			\cline{2-19}
			& \multirow{2}{*}{4} & 9 & 375 & \textbf{1,611} & - & - & \multirow{2}{*}{5} & 12 & 429 & \textbf{19,306} & - & - & \multirow{2}{*}{6} & 17 & 493 & \textbf{63,085} & - & - \\
			&  & 16 & 381 & \textbf{2,553} & - & - &  & 19 & 429 & \textbf{4,802} & - & - &  & 20 & 493 & \textbf{14,091} & - & - \\
			\hline
			\multirow{4}{*}{LKML} 
			& \multirow{2}{*}{1} & 3 & 462 & \textbf{717} & 6,515 & 3,739 & \multirow{2}{*}{2} & 7 & 94 & \textbf{604} & - & - & \multirow{2}{*}{3} & 9 & 112 & \textbf{6,472} & - & - \\
			&  & 6 & 85 & \textbf{11} & 409 & 144 &  & 9 & 85 & \textbf{2.77} & 16,231 & 15,665 &  & 11 & 107 & \textbf{1.07} & - & - \\
			\cline{2-19}
			& \multirow{2}{*}{4} & 11 & 129 & \textbf{5,713} & - & - & \multirow{2}{*}{5} & 13 & 147 & \textbf{1,434} & - & - & \multirow{2}{*}{6} & 15 & 161 & \textbf{58} & - & - \\
			&  & 12 & 129 & \textbf{47} & - & - &  & 14 & 146 & \textbf{1.70} & - & - &  & 16 & 0 & \textbf{0.15} & - & - \\
			\hline
			%\multirow{4}{*}{Team} 
			%& \multirow{2}{*}{1} & 3 & 267 & 1128 & 3963 & 7801 & \multirow{2}{*}{2} & 8 & 54 & 41 & 64 & 862 & \multirow{2}{*}{3} & 9 & 65 & 3602 & 18125 & 18223 \\
			%&  & 6 & 79 & 9.29 & 8.98 & 72 &  & 11 & 0 & 0.00 & 0.02 & 7818 &  & 12 & 0 & 0.00 & 0.02 & 7588 \\
			%\cline{2-19}
			%& \multirow{2}{*}{4} & 11 & 0 & 142 & 4208 & 7614 & \multirow{2}{*}{5} & 12 & 89 & 2646 & - & - & \multirow{2}{*}{6} & 14 & 0 & 17 & - & - \\
			%&  & 12 & 0 & 1.11 & 5.77 & 559 &  & 13 & 0 & 3.36 & - & - &  & 15 & 0 & 0.08 & - & - \\
			%\hline
			\multirow{4}{*}{Mummun} 
			& \multirow{2}{*}{1} & 3 & 8,805 & \textbf{6,638} & - & - & \multirow{2}{*}{2} & 12 & 510 & \textbf{1,889} & 9,124 & 20,947 & \multirow{2}{*}{3} & 14 & 568 & \textbf{1,146} & - & - \\
			&  & 6 & 1,783 & \textbf{2,907} & 18,273 & 11,284 &  & 15 & 510 & \textbf{2.49} & 244 & 46 &  & 18 & 551 & \textbf{0.18} & 662 & 81 \\
			\cline{2-19}
			& \multirow{2}{*}{4} & 16 & 610 & \textbf{611} & - & - & \multirow{2}{*}{5} & 18 & 670 & \textbf{214} & - & - & \multirow{2}{*}{6} & 19 & 717 & \textbf{1,761} & - & - \\
			&  & 19 & 580 & \textbf{0.33} & - & 9,257 &  & 20 & 670 & \textbf{0.30} & - & - &  & 20 & 684 & \textbf{0.39} & - & - \\
			\hline
			\multirow{4}{*}{Citeu} 
			& \multirow{2}{*}{1} & 3 & 12,429 & \textbf{92} & 24,568 & 175 & \multirow{2}{*}{2} & 11 & 2,600 & \textbf{1,774} & - & - & \multirow{2}{*}{3} & 14 & 2,600 & \textbf{4,670} & - & - \\
			&  & 6 & 4,970 & \textbf{1,898} & 23,042 & 28,228 &  & 14 & 2,600 & \textbf{2.71} & - & - &  & 17 & 2,600 & \textbf{2.13} & - & - \\
			\cline{2-19}
			& \multirow{2}{*}{4} & 16 & 2,600 & \textbf{36} & - & - & \multirow{2}{*}{5} & 17 & 2,600 & \textbf{65} & - & - & \multirow{2}{*}{6} & 18 & 2,600 & \textbf{471} & - & - \\
			&  & 20 & 2,600 & \textbf{0.82} & - & - &  & 20 & 2,600 & \textbf{2.34} & - & - &  & 20 & 2,600 & \textbf{5.17} & - & - \\
			\hline
			\multirow{4}{*}{IMDB} 
			& \multirow{2}{*}{1} & 3 & 564 & \textbf{28} & 1,164 & 528 & \multirow{2}{*}{2} & 5 & 554 & \textbf{1,943} & - & - & \multirow{2}{*}{3} & 8 & 554 & \textbf{278} & - & - \\
			&  & 6 & 529 & 5.15 & 16 & \textbf{4.39} &  & 8 & 516 & \textbf{17} & 1,635 & 618 &  & 11 & 554 & \textbf{144} & 52,220 & 18,534 \\
			\cline{2-19}
			& \multirow{2}{*}{4} & 9 & 593 & \textbf{5,723} & - & - & \multirow{2}{*}{5} & 16 & 648 & \textbf{4,100} & - & - & \multirow{2}{*}{6} & 15 & 719 & \textbf{80,574} & - & - \\
			&  & 12 & 593 & \textbf{3,967} & - & - &  & 19 & 648 & \textbf{2,594} & - & - &  & 19 & 719 & \textbf{7,645} & - & - \\
			\hline
			\multirow{4}{*}{Amazon} 
			& \multirow{2}{*}{1} & 3 & 433 & \textbf{272} & 5,150 & 2,277 & \multirow{2}{*}{2} & 5 & 461 & \textbf{248} & - & - & \multirow{2}{*}{3} & 7 & 500 & \textbf{1,508} & - & - \\
			&  & 6 & 433 & \textbf{13} & 565 & 437 &  & 8 & 461 & \textbf{11} & - & - &  & 10 & 500 & \textbf{12} & - & - \\
			\cline{2-19}
			& \multirow{2}{*}{4} & 11 & 527 & \textbf{26} & - & - & \multirow{2}{*}{5} & 12 & 553 & \textbf{680} & - & - & \multirow{2}{*}{6} & 18 & 324 & \textbf{7,752} & - & - \\
			&  & 14 & 325 & \textbf{2,937} & - & - &  & 17 & 299 & \textbf{337} & - & - &  & 19 & 324 & \textbf{64} & - & - \\
			\hline
			%\multirow{4}{*}{Twitter} 
			%& \multirow{2}{*}{1} & 16 & 1,408 & 3340 & 20624 & - & \multirow{2}{*}{2} & 5 & 5,369 & 12 & - & - & \multirow{2}{*}{3} & 7 & 5,369 & 8.77 & - & - \\
			%&  & 19 & 1,361 & 7.00 & 6.10 & - &  & 8 & 5,369 & 5.15 & - & - &  & 10 & 5,369 & 6.80 & - & - \\
			%\cline{2-19}
			%& \multirow{2}{*}{4} & 9 & 5,369 & 6.42 & - & - & \multirow{2}{*}{5} & 11 & 5,407 & 6.58 & - & - & \multirow{2}{*}{6} & 19 & 5,577 & 7.00 & - & - \\
			%&  & 12 & 5,407 & 4.93 & - & - &  & 13 & 5,385 & 4.93 & - & - &  & 20 & - & - & - & - \\
			%\hline
			\multirow{4}{*}{Aol} 
			& \multirow{2}{*}{1} & 4 & 856 & \textbf{9,516} & - & - & \multirow{2}{*}{2} & 8 & 250 & \textbf{5,095} & - & - & \multirow{2}{*}{3} & 11 & 229 & \textbf{9,230} & - & - \\
			&  & 8 & 209 & \textbf{21} & - & 72,501 &  & 11 & 194 & \textbf{88} & - & - &  & 14 & 219 & \textbf{87} & - & - \\
			\cline{2-19}
			& \multirow{2}{*}{4} & 15 & 256 & \textbf{1,779} & - & - & \multirow{2}{*}{5} & 17 & 281 & \textbf{2,195} & - & - & \multirow{2}{*}{6} & 18 & 329 & \textbf{39,285} & - & - \\
			&  & 17 & 249 & \textbf{6.14} & - & - &  & 18 & 280 & \textbf{31} & - & - &  & 19 & 320 & \textbf{949} & - & - \\
			\hline
			\multirow{4}{*}{Google} 
			& \multirow{2}{*}{1} & 6 & 204 & \textbf{2,346} & 80,534 & 18,103 & \multirow{2}{*}{2} & 9 & 193 & \textbf{6,112} & - & - & \multirow{2}{*}{3} & 13 & 209 & \textbf{825} & - & - \\
			&  & 9 & 148 & \textbf{25} & 118 & 97 &  & 12 & 178 & \textbf{13} & 961 & 618 &  & 15 & 0 & \textbf{5.66} & 4,426 & - \\
			\cline{2-19}
			& \multirow{2}{*}{4} & 15 & 233 & \textbf{617} & - & - & \multirow{2}{*}{5} & 17 & 237 & \textbf{126} & - & - & \multirow{2}{*}{6} & 18 & 258 & \textbf{3,730} & - & - \\
			&  & 17 & 0 & \textbf{2.16} & 54,390 & 30,586 &  & 18 & 0 & \textbf{5.39} & - & - &  & 20 & 0 & \textbf{1.88} & - & - \\
			\hline
		\end{tabular}
}\end{table*}

\subsection{Experimental Setup}
\noindent\textbf{Datasets.}
We use 8 real-world bipartite graphs from diverse categories to evaluate the efficiency of our algorithm. These graphs have been widely used as benchmark datasets for bipartite cohesive subgraph search problems \cite{RN13,RN71,RN66,RN152}. The detailed information of these graphs is shown in Table~\ref{tab:datasets}, where $\delta$ represents the maximum degree of vertices on each side. All these datasets are publicly available at \url{http://konect.cc/networks}.

\noindent\textbf{Algorithms.}
We implement our proposed deletion-based algorithm in Algorithm~\ref{alg:bdd}, denoted by \textsf{DMBP}, which is equipped with all optimization techniques described in Sec.~\ref{sec:ub}. We also implement two state-of-the-art maximum $k$-biplex search algorithms for comparative analysis, namely \fastbb and \cpc. Specifically, \fastbb is a branch-and-bound algorithm that adopts the symmetric-BK branching strategy, which achieves a non-trivial worst-case time complexity \cite{RN71}. \cpc is a core-based algorithm that combines the symmetric-BK branching strategy with a core-based pruning method \cite{RN152}. 
All algorithms are implemented in \texttt{C++} and executed on a PC equipped with a \texttt{2.2GHz} CPU and \texttt{128GB} RAM running \texttt{Linux}.

\noindent\textbf{Parameter settings.}
During the evaluations of all algorithms, we vary $k$ from $1$ to $6$. Following the size constraint in Sec.~2, we choose $\theta$ from the range $[2k+1,20]$ for each $k$. We report the CPU time in seconds, excluding the time for reading graphs from external storage. Any runtime longer than $24$ hours is regarded as a timeout and denoted by ``-'' in experimental results.

\begin{figure*}[!t]
	\centering
	\includegraphics[width=\textwidth]{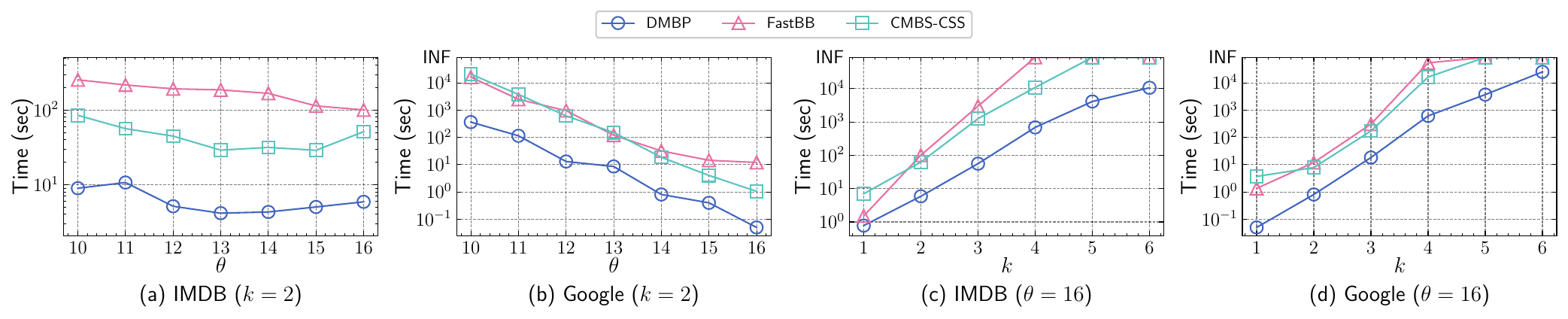}
	\caption{Runtime of different algorithms with varying $\theta$ and $k$}
	\label{fig:varying-parameters}
\end{figure*}

\subsection{Experimental Results}

\noindent\textbf{Exp-1: Efficiency comparison among different algorithms.}
We evaluate the efficiency of \textsf{DMBP} and two baseline algorithms, \fastbb and \cpc, on 8 real-world graphs. The results are shown in Table~\ref{tab:overall-runtime}, where the column $\lvert E^*\rvert$ represents the number of edges in the maximum $k$-biplex, and the notion ``-'' represents a timeout exceeding 24 hours. A value $|E^*|=0$ means that no $k$-biplex satisfying the size threshold exists for this parameter setting. As observed, \textsf{DMBP} consistently outperforms the baseline algorithms across most input parameters. Specifically, \textsf{DMBP} solves all 96 test cases, while \fastbb and \cpc solve only 23 and 22 test cases, respectively. The advantage becomes more evident when $k$ is large: among the 64 test cases with $k\ge3$, \fastbb and \cpc solve only 4 and 2 test cases, respectively. For example, on \textit{LKML} with $k=2$ and $\theta=9$, \textsf{DMBP} completes in $2.77$s, while \fastbb and \cpc take $16,231$s and $15,665$s, respectively, achieving at least $5000\times$ speedups. On \textit{Mummun} with $k=4$ and $\theta=19$, \textsf{DMBP} completes in $0.33$s, while \cpc takes $9,257$s and \fastbb times out, achieving at least $28000\times$ speedups. Notably, \textsf{DMBP} solves all test cases on \textit{Aol} with varying $k$ from $1$ to $6$. In contrast, the two baseline algorithms fail to solve any test cases on this graph when $k\ge 2$, and \cpc barely managed to finish when $k=1$ and $\theta=8$ in 21 hours. These results demonstrate the high efficiency of our proposed algorithm, especially with large $k$ values.

\noindent\textbf{Exp-2: Efficiency with varying $\theta$ and $k$.}
We evaluate the performance of \dmbp and the two baseline algorithms \fastbb and \cpc by varying $\theta$ and $k$ on two representative datasets, \textit{IMDB} and \textit{Google}. Similar results can be obtained on other datasets. The results are shown in Fig.~\ref{fig:varying-parameters}, where the notion ``INF'' represents a timeout. As observed, \textsf{DMBP} consistently outperforms the baselines under different parameter settings, achieving speedups up to $100\times$. For example, on the graph \textit{IMDB} with $k=4$ and $\theta=16$ shown in Fig.~\ref{fig:varying-parameters}(c), our algorithm takes only $80$ seconds, while \cpc takes over $1000$ seconds and \fastbb cannot finish in 24 hours. When $\theta$ decreases or $k$ increases, the search space becomes larger and all algorithms require more time. However, the runtime curve of \textsf{DMBP} grows much more slowly than those of the baselines, while \fastbb and \cpc frequently reach the timeout on hard $k$ or $\theta$ settings. These results indicate that \textsf{DMBP} is less sensitive to input parameters, benefiting from the deletion-based branching strategy in Sec.~\ref{sec:dbs} and the proposed upper-bounding techniques in Sec.~\ref{sec:ub}.

\begin{figure*}[!t]
	\centering
	\includegraphics[width=\textwidth]{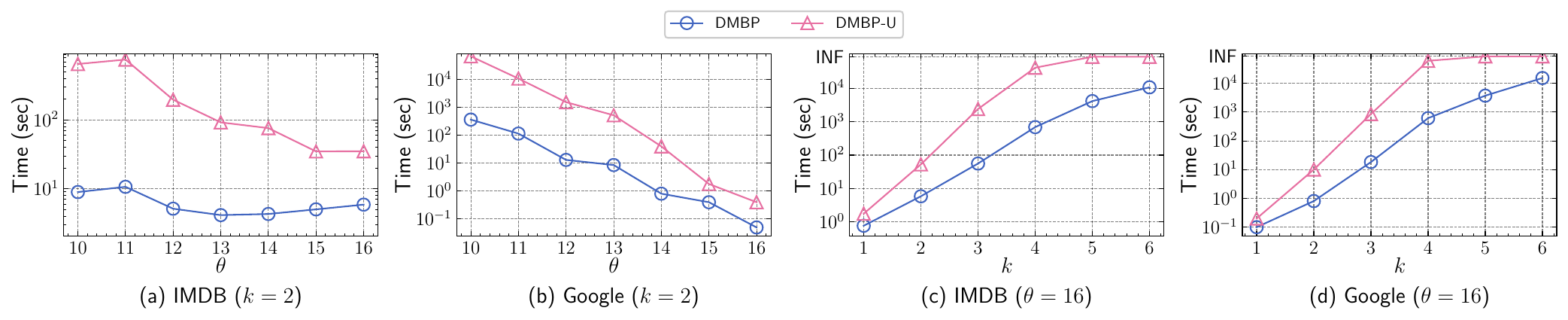}
	\caption{Runtime without using the proposed upper-bounding techniques}
	\label{fig:upper-bound}
\end{figure*}

\begin{figure*}[!t]
	\centering
	\includegraphics[width=\textwidth]{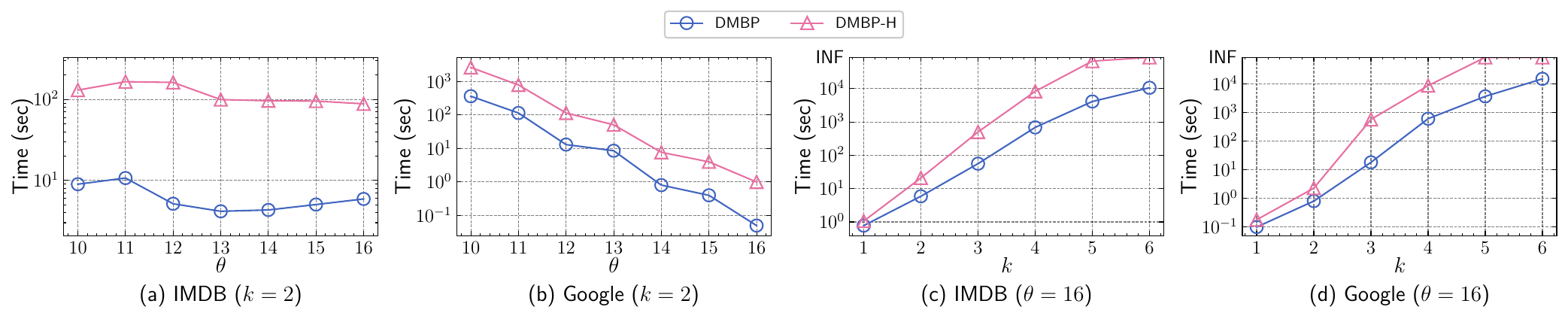}
	\caption{Runtime without using the heuristic algorithm}
	\label{fig:heuristic-runtime}
\end{figure*}

\begin{figure*}[!t]
	\centering
	\includegraphics[width=\textwidth]{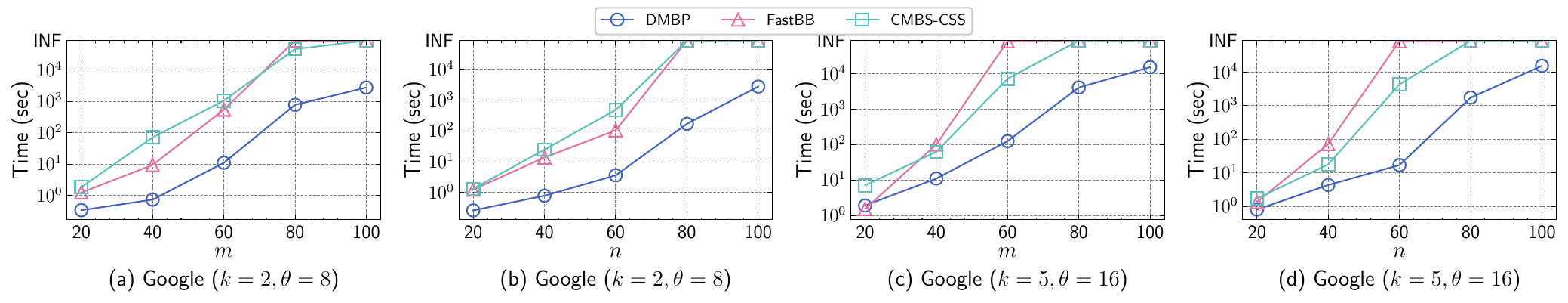}
	\caption{Scalability of different algorithms under vertex and edge sampling}
	\label{fig:scalability}
\end{figure*}

\noindent\textbf{Exp-3: Efficiency of upper bounding techniques.}
We evaluate the effectiveness of the proposed upper-bounding techniques by comparing \textsf{DMBP} with \textsf{DMBP-U}, which removes the upper bounds proposed in Sec.~\ref{sec:ub} from our deletion-based algorithm. The results are shown in Fig.~\ref{fig:upper-bound}. As observed, incorporating the upper-bounding techniques significantly improves the efficiency of \textsf{DMBP}. The improvement becomes more pronounced as $k$ increases or $\theta$ decreases, since these settings generate more partial solutions and make upper-bound pruning more important. In particular, \textsf{DMBP-U} can be up to two orders of magnitude slower than \textsf{DMBP} and may even time out on hard settings, while \textsf{DMBP} still finishes all the test cases efficiently. For example, on the graph \textit{Google} with $k=4$ and $\theta=16$, \textsf{DMBP} takes about 80 seconds while \textsf{DMBP-U} takes more than 22 hours. These results demonstrate the strong pruning power of the proposed upper bounds.

\noindent\textbf{Exp-4: Efficiency of the heuristic algorithm.}
We evaluate the effectiveness of the heuristic algorithm by comparing \textsf{DMBP} with \textsf{DMBP-H}, which starts from an empty initial solution instead of the solution returned by Algorithm~\ref{alg:heu}. The results are shown in Fig.~\ref{fig:heuristic-runtime}. As observed, \textsf{DMBP} consistently outperforms \textsf{DMBP-H}. This is because the heuristic algorithm provides a large $k$-biplex at the beginning, which tightens the pruning condition provided by the upper-bounding techniques in Sec.~\ref{sec:ub}. The improvement of our heuristic algorithm is quite stable when varying the value of $\theta$, but becomes more significant on difficult settings with a larger $k$, where the total number of branches grows exponentially and a strong initial solution helps prune many redundant branches. These results show that the heuristic algorithm can substantially reduce the search cost by providing a strong initial solution.

\noindent\textbf{Exp-5: Solution quality of the heuristic algorithm.}
We further verify the effectiveness of our heuristic algorithm by comparing the initial solution returned by the heuristic algorithm with the exact maximum $k$-biplex obtained by \textsf{DMBP}. The results are reported in Table~\ref{tab:heu}, where $\lvert E_H\rvert$ and $\lvert E^*\rvert$ denote the numbers of edges in the heuristic solution and the maximum $k$-biplex, respectively. Table~\ref{tab:heu} shows that the heuristic algorithm obtains solutions close to the optimum on all tested cases. The edge ratio $\lvert E_H\rvert/\lvert E^*\rvert$ ranges from $85.2\%$ to $98.1\%$, with an average value of $92.9\%$. For example, on \textit{IMDB} with $k=4$ and $\theta=20$, the heuristic algorithm returns a solution with 563 edges, while the optimal solution contains 574 edges. When comparing the running time, the heuristic algorithm is always significantly faster than \textsf{DMBP}. For example, the heuristic algorithm takes only $0.51$ seconds on \textit{IMDB} with $k=4$ and $\theta=20$, whereas the exact search takes $190$ seconds. These results confirm that the heuristic algorithm can efficiently provide high-quality initial solutions for \textsf{DMBP}.

\begin{table}
	\small
	\caption{Solution quality of the heuristic algorithm}
	\label{tab:heu}
	\begin{tabular}{c|c|c|c|c|c|c}
		\hline
		Dataset&$k$&$\theta$&$\lvert E_H\rvert$&$\lvert E^*\rvert$&\heu time&\dmbp time\\
		\hline
		\multirow{3}{*}{IMDB}&2&13&483&516&0.64&4.15 \\
		&3&17&501&522&0.39&56\\
		&4&20&563&574&0.51&190\\
		\hline
		\multirow{3}{*}{Google}&2&10&156&183&10&365 \\
		&3&13&198&209&6.53&825\\
		&4&15&209&233&4.98&617\\
		\hline
		
	\end{tabular}
\end{table}

\noindent\textbf{Exp-6: Scalability testing.}
We evaluate the scalability of algorithms on the \textit{Google} dataset using random sampling of vertices and edges. Results on the other datasets show similar trends. We prepare sampled subgraphs by sampling 20\%, 40\%, 60\%, and 80\% vertices or edges from the original graph. The results are shown in Fig.~\ref{fig:scalability}. As observed, \textsf{DMBP} shows the slowest runtime growth as the graph size increases, while the baseline algorithms exhibit much steeper growth. On the 60\% vertex-sampled graph with $k=5$ and $\theta=16$, \textsf{DMBP} completes in $20$ seconds, whereas \cpc takes more than one hour and \fastbb times out. These results indicate that the proposed deletion-based framework achieves the best scalability among the tested algorithms.

\begin{figure}[!t]
	\centering
	\includegraphics[width=0.9\columnwidth]{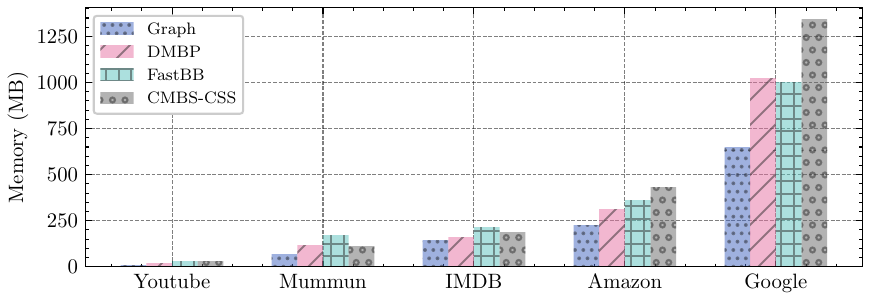}
	\caption{Memory usage of different algorithms}
	\label{fig:memory}
\end{figure}

\noindent\textbf{Exp-7: Memory usage testing.}
We evaluate the physical memory usage of \textsf{DMBP}, \fastbb, and \cpc on 5 representative datasets. Similar results can be obtained on other graphs. The results are shown in Fig.~\ref{fig:memory}. As observed, the memory usage of all algorithms grows almost linearly with the graph size. \textsf{DMBP} consumes memory comparable to \fastbb, and both of them require less memory than \cpc. This is because the core-based algorithm relies on maintaining a large number of cores to reduce the graph size, which introduces additional memory overhead. These results indicate that \textsf{DMBP} is space efficient on real-world graphs.

\vspace{5pt}

\section{Related Works}

\noindent{\textbf{Maximal $k$-biplex enumeration.}} Closely related to our work is the maximal $k$-biplex enumeration problem, which aims to find all $k$-biplexes not contained in any larger ones. This problem has attracted increasing attention in recent years \cite{RN72,RN74,RN75,RN66,RN151}. Sim et al. \cite{RN151} first defined the $k$-biplex as a type of quasi-biclique and proposed a basic branch-and-bound algorithm \textit{MQBminer} to enumerate all maximal $k$-biplexes satisfying a vertex size constraint. Although it uses degree-based and common-neighbor-based pruning methods, \textit{MQBminer} still generates many redundant branches.
To address this, Yu et al. \cite{RN74} proposed \textit{iMB}, a prefix-based algorithm equipped with early stopping strategies to reduce redundant non-maximal branches. However, its performance degrades on graphs with millions of edges when $k > 2$. Yu et al. \cite{RN74} also introduced \textit{iTraversal}, a polynomial-delay algorithm based on a reverse search framework. While it guarantees polynomial time between outputs, it requires storing all maximal $k$-biplexes in memory, making it unsuitable for large-scale graphs.
Alternatively, Dai et al. \cite{RN66} proposed a pivot-based algorithm that uses pivoting techniques to avoid numerous non-maximal branches. However, it still suffers from a trivial worst-case time complexity of $O^*(2^n)$. To overcome this, Dai et al. \cite{RN75} later developed two algorithms combining pivoting techniques with a binary branching strategy. These two algorithms achieve non-trivial time complexities and currently offer the best practical performance when $k\le 6$. These algorithms can be adapted to maximum $k$-biplex search by enumerating all maximal $k$-biplexes and returning the one with the most edges. However, since the number of maximal $k$-biplexes can be exponential in the graph size, this approach may enumerate many maximal but non-maximum $k$-biplexes, causing substantial unnecessary computation in practice.

\noindent{\textbf{Maximum $k$-plex search.}} The maximum $k$-plex search problem aims to find a $k$-plex with the most vertices in traditional graphs, where the $k$-plex model was originally introduced in \cite{RN148}. This domain encompasses various formulations, among which the maximum $k$-plex search problem is the closest to our work and has been proven to be NP-hard \cite{RN149}. Many approaches have been proposed to address this challenge. On the one hand, some approaches focus on carefully designed branching rules to surpass the trivial $O^*(2^n)$ barrier, including strategies that prioritize vertices based on non-neighbor counts to strictly limit worst-case complexity \cite{RN137,RN141}. Notably, Wang et al. \cite{RN140} developed a fixed-parameter tractable algorithm by exploiting the dual problem of the maximum $k$-plex search and derived a worst-case time complexity of $O^*((k+1)^g)$, where $g$ represents the gap between the degeneracy and the size of maximum $k$-plex. On the other hand, some approaches target scalability on large-scale graphs by using graph reductions based on degree and common-neighbor constraints \cite{RN138,RN142,RN144}, as well as new upper bounds derived from vertex coloring and vertex partitioning \cite{RN144,RN145,RN146,RN150}, to prune unnecessary branches. Notably, Ahmad et al. \cite{RN139} synthesized these state-of-the-art techniques into a comprehensive framework to identify optimal configurations for general scenarios. However, these approaches are designed to maximize the number of vertices in traditional graphs, which is quite different from our problem of maximizing the number of edges in bipartite graphs, so these methods cannot be directly applied to the maximum $k$-biplex search problem.

%\noindent{\textbf{Maximum biclique search.}} The maximum biclique search problem is widely categorized into three variants: Maximum Edge Biclique (MEB), Maximum Vertex Biclique (MVB), and Maximum Balanced Biclique (MBB). The MEB problem aims to maximize the number of edges, which is proven to be NP-hard \cite{RN17} and inapproximable \cite{RN79}. Various approaches have been proposed to address it, including a probabilistic algorithm based on Monte-Carlo subspace clustering \cite{RN19}, an integer programming method with several scale reductions \cite{RN20}, and an advanced branch-and-bound frameworks capable of handling billion-scale graphs via progressive bounding \cite{RN13}. Recently, Dai et al. \cite{RN162} proposed a hybrid algorithm that combines the pivoting technique and the minimal vertex cover search, which achieves the best time complexity of $O^*(1.381^n)$. The MVB problem aims to find a biclique that maximizes the total vertex count. Unlike MEB, this variant can be solved in polynomial time via minimum-cut algorithms \cite{RN36}. The MBB problem, which requires equal-sized vertex sets, remains computationally challenging. Approaches to this problem include evolutionary algorithms with structure mutation \cite{RN28}, heuristic-driven local search frameworks \cite{RN24}, and exact algorithms utilizing upper bound propagation \cite{RN32}. Notably, Chen et al. \cite{RN23} recently optimized branch-and-bound techniques using bipartite sparsity measurements are proposed to improve performance on both small dense and large sparse graphs.

\section{Conclusion}
In this paper, we investigate the maximum $k$-biplex search problem in bipartite graphs. We propose a novel deletion-based algorithm that reformulates the problem as finding a minimal $k$-bounded degree deletion in the complement graph. Based on this formulation, \textsf{DMBP} employs a new deletion-based branching strategy and branch reduction rules, achieving a worst-case time complexity below $O^*(2^n)$. In addition, we develop several optimization techniques to further improve practical efficiency, including effective upper bounds and a heuristic algorithm. Extensive experiments on 8 real-world bipartite graphs demonstrate the efficiency and scalability of our deletion-based algorithm and the proposed optimization techniques.

\bibliographystyle{ACM-Reference-Format}
\balance\bibliography{refs}

@inproceedings{RN72,
	author = {Yu, Kaiqiang and Long, Cheng and Liu, Shengxin and Yan, Da},
	title = {Efficient Algorithms for Maximal k-Biplex Enumeration},
	year = {2022},
	booktitle = {Proceedings of the 2022 International Conference on Management of Data},
	pages = {860–873},
	numpages = {14},
}

@article{RN58,
   author = {Alzahrani, Taher and Horadam, Kathy},
   title = {Finding maximal bicliques in bipartite networks using node similarity},
   journal = {Applied Network Science},
   volume = {4},
   number = {1},
   pages = {21},
   ISSN = {2364-8228},
   year = {2019},
   type = {Journal Article}
}

@article{RN78,
   author = {Batagelj, Vladimir and Zaversnik, Matjaz},
   title = {An o (m) algorithm for cores decomposition of networks},
   journal = {arXiv preprint cs/0310049},
   year = {2003},
   type = {Journal Article}
}

@inproceedings{RN68,
   author = {Braun, Fabian and Caelen, Olivier and Smirnov, Evgueni N. and Kelk, Steven and Lebichot, Bertrand},
   title = {Improving Card Fraud Detection Through Suspicious Pattern Discovery},
   booktitle = {Advances in Artificial Intelligence: From Theory to Practice},
   editor = {Benferhat, Salem and Tabia, Karim and Ali, Moonis},
   publisher = {Springer International Publishing},
   pages = {181-190},
   ISBN = {978-3-319-60045-1},
   year = {2017},
   type = {Conference Proceedings}
}

@article{RN66,
   author = {Dai, Qiangqiang and Li, Rong-Hua and Ye, Xiaowei and Liao, Meihao and Zhang, Weipeng and Wang, Guoren},
   title = {Hereditary Cohesive Subgraphs Enumeration on Bipartite Graphs: The Power of Pivot-based Approaches},
   journal = {Proceedings of the ACM on Management of Data},
   volume = {1},
   number = {2},
   pages = {1-26},
   ISSN = {2836-6573},
   year = {2023},
   type = {Journal Article}
}

@article{RN75,
   author = {Dai, Qiangqiang and Li, Rong-Hua and Cui, Donghang and Liao, Meihao and Qiu, Yu-Xuan and Wang, Guoren},
   title = {Efficient Maximal Biplex Enumerations with Improved Worst-Case Time Guarantee},
   journal = {Proceedings of the ACM on Management of Data},
   volume = {2},
   number = {3},
   pages = {1-26},
   ISSN = {2836-6573},
   year = {2024},
   type = {Journal Article}
}

@article{RN38,
   author = {Eren, Kemal and Deveci, Mehmet and Küçüktunç, Onur and Çatalyürek, Ümit V},
   title = {A comparative analysis of biclustering algorithms for gene expression data},
   journal = {Briefings in bioinformatics},
   volume = {14},
   number = {3},
   pages = {279-292},
   ISSN = {1477-4054},
   year = {2013},
   type = {Journal Article}
}

@article{RN22,
   author = {Feng, Qilong and Li, Shaohua and Zhou, Zeyang and Wang, Jianxin},
   title = {Parameterized algorithms for edge biclique and related problems},
   journal = {Theoretical Computer Science},
   volume = {734},
   pages = {105-118},
   ISSN = {0304-3975},
   year = {2018},
   type = {Journal Article}
}

@article{RN14,
   author = {Gillis, Nicolas and Glineur, François},
   title = {A continuous characterization of the maximum-edge biclique problem},
   journal = {Journal of Global Optimization},
   volume = {58},
   number = {3},
   pages = {439-464},
   ISSN = {0925-5001},
   year = {2014},
   type = {Journal Article}
}

@inproceedings{RN41,
   author = {Gunnemann, Stephan and Muller, Emmanuel and Raubach, Sebastian and Seidl, Thomas},
   title = {Flexible fault tolerant subspace clustering for data with missing values},
   booktitle = {2011 IEEE 11th International Conference on Data Mining},
   publisher = {IEEE},
   pages = {231-240},
   ISBN = {1457720752},
   year = {2011},
   type = {Conference Proceedings}
}

@inproceedings{RN69,
   author = {Langston, MA and Chesler, Elissa J and Zhang, Y},
   title = {On finding bicliques in bipartite graphs: a novel algorithm with application to the integration of diverse biological data types},
   booktitle = {Proceedings of the 41st Annual Hawaii International Conference on System Sciences (HICSS 2008)(HICSS)},
   volume = {1},
   pages = {473},
   year = {2008},
   type = {Conference Proceedings}
}

@article{RN57,
   author = {Lehmann, Sune and Schwartz, Martin and Hansen, Lars Kai},
   title = {Biclique communities},
   journal = {Physical Review E—Statistical, Nonlinear, and Soft Matter Physics},
   volume = {78},
   number = {1},
   pages = {016108},
   ISSN = {1550-2376},
   year = {2008},
   type = {Journal Article}
}

@inproceedings{RN34,
   author = {Ley, Michael},
   title = {The DBLP Computer Science Bibliography: Evolution, Research Issues, Perspectives},
   series = {String Processing and Information Retrieval},
   publisher = {Springer Berlin Heidelberg},
   pages = {1-10},
   ISBN = {978-3-540-45735-0},
   year = {2002},
   type = {Conference Proceedings}
}

@misc{RN50,
   author = {Li, Jingdong and Li, Zhao and Wang, Xiaoling and Lu, Xingjian and Zhang, Ji and Chen, Hongyang},
   title = {GPU-Accelerated Maximal Bicliques Mining Framework for Large E-commerce Networks},
   pages = {539-544},
   year = {2023},
   type = {Conference Paper}
}

@inproceedings{RN70,
   author = {Li, Jinyan and Sim, Kelvin and Liu, Guimei and Wong, Limsoon},
   title = {Maximal quasi-bicliques with balanced noise tolerance: Concepts and co-clustering applications},
   booktitle = {Proceedings of the 2008 SIAM International Conference on Data Mining},
   publisher = {SIAM},
   pages = {72-83},
   year = {2008},
   type = {Conference Proceedings}
}

@inproceedings{RN47,
   author = {Liu, Guimei and Sim, Kelvin and Li, Jinyan},
   title = {Efficient mining of large maximal bicliques},
   booktitle = {Data Warehousing and Knowledge Discovery: 8th International Conference, DaWaK 2006, Krakow, Poland, September 4-8, 2006. Proceedings 8},
   publisher = {Springer},
   pages = {437-448},
   ISBN = {3540377360},
   year = {2006},
   type = {Conference Proceedings}
}

@article{RN13,
   author = {Lyu, Bingqing and Qin, Lu and Lin, Xuemin and Zhang, Ying and Qian, Zhengping and Zhou, Jingren},
   title = {Maximum biclique search at billion scale},
   journal = {Proceedings of the VLDB Endowment},
   ISSN = {2150-8097},
   year = {2020},
   type = {Journal Article}
}

@inproceedings{RN42,
   author = {Poernomo, Ardian Kristanto and Gopalkrishnan, Vivekanand},
   title = {Towards efficient mining of proportional fault-tolerant frequent itemsets},
   booktitle = {Proceedings of the 15th ACM SIGKDD international conference on Knowledge discovery and data mining},
   pages = {697-706},
   year = {2009},
   type = {Conference Proceedings}
}

@article{RN60,
   author = {Voggenreiter, Oliver and Bleuler, Stefan and Gruissem, Wilhelm},
   title = {Exact biclustering algorithm for the analysis of large gene expression data sets},
   journal = {BMC bioinformatics},
   volume = {13},
   number = {Suppl 18},
   pages = {A10},
   ISSN = {1471-2105},
   year = {2012},
   type = {Journal Article}
}

@article{RN40,
   author = {Xie, Juan and Ma, Anjun and Fennell, Anne and Ma, Qin and Zhao, Jing},
   title = {It is time to apply biclustering: a comprehensive review of biclustering applications in biological and biomedical data},
   journal = {Briefings in bioinformatics},
   volume = {20},
   number = {4},
   pages = {1450-1465},
   ISSN = {1467-5463},
   year = {2019},
   type = {Journal Article}
}

@article{RN71,
   author = {Yu, Kaiqiang and Long, Cheng},
   title = {Maximum k-Biplex Search on Bipartite Graphs: A Symmetric-BK Branching Approach},
   journal = {Proc. ACM Manag. Data},
   volume = {1},
   number = {1},
   pages = {Article 49},
   year = {2023},
   type = {Journal Article}
}

@inproceedings{RN73,
	author = {Luo, Wensheng and Li, Kenli and Zhou, Xu and Gao, Yunjun and Li, Keqin},
	title = {Maximum biplex search over bipartite graphs},
	booktitle = {2022 IEEE 38th International Conference on Data Engineering (ICDE)},
	publisher = {IEEE},
	pages = {898-910},
	ISBN = {1665408839},
	type = {Conference Proceedings}
}

@inproceedings{RN130,
	author = {Liu, Xiaowen and Li, Jinyan and Wang, Lusheng},
	title = {Quasi-bicliques: Complexity and Binding Pairs},
	series = {Computing and Combinatorics},
	publisher = {Springer Berlin Heidelberg},
	pages = {255-264},
	ISBN = {978-3-540-69733-6},
	type = {Conference Proceedings}
}

@article{RN137,
	author = {Chang, Lijun and Yao, Kai},
	title = {Maximum k-plex computation: Theory and practice},
	journal = {Proceedings of the ACM on Management of Data},
	volume = {2},
	number = {1},
	pages = {1-26},
	ISSN = {2836-6573},
	year = {2024},
	type = {Journal Article}
}

@article{RN138,
	author = {Chang, Lijun and Xu, Mouyi and Strash, Darren},
	title = {Efficient maximum k-plex computation over large sparse graphs},
	journal = {Proceedings of the VLDB Endowment},
	volume = {16},
	number = {2},
	pages = {127-139},
	ISSN = {2150-8097},
	year = {2022},
	type = {Journal Article}
}

@article{RN139,
	author = {Ahmad, Akhlaque and Yan, Da and Chen, Xiao and Yuan, Lyuheng and Zhang, Qin and Adhikari, Saugat},
	title = {Maximum k-Plex Finding: Choices of Pruning Techniques Matter!},
	journal = {Proceedings of the VLDB Endowment},
	volume = {18},
	number = {9},
	pages = {2928-2940},
	ISSN = {2150-8097},
	year = {2025},
	type = {Journal Article}
}

@inproceedings{RN140,
	author = {Wang, Zhengren and Zhou, Yi and Luo, Chunyu and Xiao, Mingyu},
	title = {A Fast Maximum k-Plex Algorithm Parameterized by the Degeneracy Gap},
	booktitle = {IJCAI},
	pages = {5648-5656},
	type = {Conference Proceedings},
	year = {2023}
}

@inproceedings{RN141,
	author = {Xiao, Mingyu and Lin, Weibo and Dai, Yuanshun and Zeng, Yifeng},
	title = {A fast algorithm to compute maximum k-plexes in social network analysis},
	booktitle = {Proceedings of the AAAI conference on Artificial Intelligence},
	volume = {31},
	ISBN = {2374-3468},
	type = {Conference Proceedings}
}

@inproceedings{RN142,
	author = {Gao, Jian and Chen, Jiejiang and Yin, Minghao and Chen, Rong and Wang, Yiyuan},
	title = {An exact algorithm for maximum k-plexes in massive graphs},
	booktitle = {IJCAI},
	pages = {1449-1455},
	type = {Conference Proceedings}
}

@inproceedings{RN144,
	author = {Zhou, Yi and Hu, Shan and Xiao, Mingyu and Fu, Zhang-Hua},
	title = {Improving maximum k-plex solver via second-order reduction and graph color bounding},
	booktitle = {Proceedings of the AAAI Conference on Artificial Intelligence},
	volume = {35},
	pages = {12453-12460},
	ISBN = {2374-3468},
	type = {Conference Proceedings}
}

@inproceedings{RN145,
	author = {Jiang, Hua and Zhu, Dongming and Xie, Zhichao and Yao, Shaowen and Fu, Zhang-Hua},
	title = {A New Upper Bound Based on Vertex Partitioning for the Maximum K-plex Problem},
	booktitle = {IJCAI},
	pages = {1689-1696},
	type = {Conference Proceedings},
	year = {2021}
}

@inproceedings{RN146,
	author = {Jiang, Hua and Xu, Fusheng and Zheng, Zhifei and Wang, Bowen and Zhou, Wei},
	title = {A Refined Upper Bound and Inprocessing for the Maximum K-plex Problem},
	booktitle = {IJCAI},
	pages = {5613-5621},
	type = {Conference Proceedings},
	year = {2023}
}

@article{RN148,
	author = {Seidman, Stephen B and Foster, Brian L},
	title = {A graph‐theoretic generalization of the clique concept},
	journal = {Journal of Mathematical sociology},
	volume = {6},
	number = {1},
	pages = {139-154},
	ISSN = {0022-250X},
	year = {1978},
	type = {Journal Article}
}

@article{RN149,
	author = {Lewis, John M. and Yannakakis, Mihalis},
	title = {The node-deletion problem for hereditary properties is NP-complete},
	journal = {Journal of Computer and System Sciences},
	volume = {20},
	number = {2},
	pages = {219-230},
	ISSN = {0022-0000},
	DOI = {https://doi.org/10.1016/0022-0000(80)90060-4},
	url = {https://www.sciencedirect.com/science/article/pii/0022000080900604},
	year = {1980},
	type = {Journal Article}
}

@inproceedings{RN150,
	author = {Zheng, Jiongzhi and Jin, Mingming and He, Kun},
	title = {Exact algorithms with new upper bounds for the maximum k-plex problem},
	booktitle = {Proceedings of the Thirty-Fourth International Joint Conference on Artificial Intelligence},
	pages = {9014-9021},
	type = {Conference Proceedings}
}

@article{RN74,
	author = {Yu, Kaiqiang and Long, Cheng and Deepak, P and Chakraborty, Tanmoy},
	title = {On efficient large maximal biplex discovery},
	journal = {IEEE Transactions on Knowledge and Data Engineering},
	volume = {35},
	number = {1},
	pages = {824-829},
	ISSN = {1041-4347},
	year = {2021},
	type = {Journal Article}
}

@article{RN151,
	author = {Sim, Kelvin and Li, Jinyan and Gopalkrishnan, Vivekanand and Liu, Guimei},
	title = {Mining maximal quasi-bicliques: Novel algorithm and applications in the stock market and protein networks},
	journal = {Stat. Anal. Data Min.},
	volume = {2},
	number = {4},
	pages = {255–273},
	ISSN = {1932-1864},
	year = {2009},
	type = {Journal Article}
}

@article{RN152,
	author = {Pan, Dong and Zhou, Xu and Luo, Wensheng and Yang, Zhibang and Liu, Qing and Gao, Yunjun and Li, Kenli},
	title = {Accelerating maximum biplex search over large bipartite graphs},
	journal = {The VLDB Journal},
	volume = {34},
	number = {1},
	pages = {1},
	ISSN = {0949-877X},
	DOI = {10.1007/s00778-024-00882-9},
	url = {https://doi.org/10.1007/s00778-024-00882-9},
	year = {2024},
	type = {Journal Article}
}

@misc{RN161,
	author = {Hao, Yang and Zhang, Mengqi and Wang, Xiaoyang and Chen, Chen},
	title = {Cohesive Subgraph Detection in Large Bipartite Networks},
	publisher = {Association for Computing Machinery},
	pages = {Article 22},
	DOI = {10.1145/3400903.3400925},
	url = {https://doi.org/10.1145/3400903.3400925},
	year = {2020},
	type = {Conference Paper}
}
	
	\let\baselinestretch\savedbaselinestretch
	
\end{document}